\documentclass[12pt,thmsa]{article}
\usepackage{sw20lart}

\input tcilatex
\begin{document}

\textbf{Exactly-solvable Ising-Heisenberg model for the coupled barotropic
fluid - rotating solid sphere system - condensation of super and
sub-rotating barotropic flow states}

By Chjan C. Lim, Mathematical Sciences, RPI, Troy, NY 12180, USA

http://www.rpi.edu/\symbol{126}limc

June 15 2006

\medskip\ 

Abstract

\smallskip\ 

Exact solutions of a family of Heisenberg-Ising spin-lattice models for a
coupled barotropic flow - massive rotating sphere system under
microcanonical constraint on relative enstrophy is obtained by the method of
spherical constraint. Phase transitions representative of Bose-Einstein
condensation in which highly ordered super and sub-rotating states
self-organize from random initial vorticity states are calculated exactly
and related to three key parameters - spin of sphere, kinetic energy of the
barotropic flow which is specified by the inverse temperature and amount of
relative enstrophy which is held fixed. Angular momentum of the barotropic
fluid relative to the rotating frame of the infinitely massive sphere is the
main order parameter in this statistical mechanics problem $-$ it is not
constrained either canonically nor microcanonically as coupling between the
fluid and the rotating sphere by a complex torque is responsible for its
change. This coupling and exchange of angular momentum is a necessary
condition for condensation in this spin-lattice system. There is no low
temperature defects in this model - the partition function is calculated in
closed form for all positive and negative temperatures. Also note-worthy is
the fact that this statistical equilibrium model is not a mean field model
and can be extended to treat fluctuations if required in more complex
coupled flows.

\section{Introduction}

Consider the system consisting of a rotating high density rigid sphere of
radius $R,$ enveloped by a thin shell of barotropic (non-divergent) fluid.
The barotropic flow is assumed to be inviscid, apart from an ability to
exchange angular momentum and energy with the heavy solid sphere. In
addition we assume that the fluid is in radiation balance and there is no
net energy gain or loss from insolation. This provides a crude model of the
complex planet - atmosphere interactions, including the enigmatic torque
mechanism responsible for the phenomenon of atmospheric super-rotation.

We will build an equilibrium statistical mechanics model to study the
relaxation of this complex phenomenon, possibly in terms of phase
transitions that are dependent on a few key parameters in the problem. For a
problem concerning super-rotation on a spherical surface there is little
doubt that one of the key parameters is angular momentum of the fluid. The
total angular momentum of the fluid and solid sphere is a conserved
quantity. We could either model the problem as such, that is, the angular
momentum of the sphere is large but finite and thus can vary, or consider
the sphere to have infinite angular momentum, in which case, it serves as a
reservoir of angular momentum and the active part of the model is just the
fluid.

It is clear that a quasi - 2d geophysical relaxation problem will involve
energy and enstrophy. The total energy of the fluid and sphere is conserved
in any frame, both rotating and inertial ones. It consists entirely of the
kinetic energy of barotropic flow plus that of the solid sphere because we
have assumed the sphere to be a rigid solid that does not deform, and there
is no gravitational potential energy in the fluid since it has uniform
thickness and density, and its upper surface is a rigid lid. Conservation of
relative enstrophy is treated here as a microcanonical constraint, modifying
the classical energy-enstrophy theories \cite{Fred} in substantial ways,
chief amongst them being removal of the Gaussian low temperature defect
while retaining the exact solvability of the model.

Higher vorticity moments are considered to be less significant than
enstrophy in statistical equilibrium models of quasi-2d geophysical flows 
\cite{Fred}. A detailed variational analysis of this topic is available in 
\cite{limshi05}.

The results in this paper is motivated by the variational analysis reported
in \cite{var} and the careful and detailed simulation results first reported
in the paper \cite{dinglim} and later in the book by Lim and Nebus, \cite
{LimNebus}. They agree in large part also with the mean field theories
reported in \cite{lim05a} and \cite{lim05b}.

\section{Energy and angular momentum in the fluid-sphere system}

The time dependent kinetic energy in the inertial frame ($\Omega =0)$ of the
barotropic fluid component of the above two components system is given by 
\[
H_0[q]=\frac 12\int_{S^2}dx\text{ }\left[ u^2+v^2\right] 
\]
where $u$ and $v$ are the zonal and meridional components of the fluid
velocity in the inertial frame.

This kinetic energy of the fluid component in the inertial frame can also be
written in terms of an arbitrary rotating frame as

\begin{eqnarray*}
H[q] &=&\frac 12\int_{S^2}dx\left[ (u_r+u_p)^2+v_r^2\right] \\
&=&\frac 12\int_{S^2}dx\left[ (u_r^2+v_r^2)+2u_ru_p\right] +\frac
12\int_{S^2}dx\text{ }u_p^2 \\
&=&-\frac 12\int_{S^2}dx\text{ }\psi q+\frac 12\int_{S^2}dx\text{ }u_p^2
\end{eqnarray*}
where $u_r=u-u_p$ is the relative zonal velocity, $\psi $ is the stream
function for the relative flow and 
\begin{eqnarray*}
q &=&\omega +2\Omega \cos \theta \\
&=&\Delta \psi +2\Omega \cos \theta
\end{eqnarray*}
is the vorticity in the rest frame in terms of the relative vorticity $%
\omega $ and the planetary vorticity $2\Omega \cos \theta $; here $\theta $
denotes co-latitude on the unit sphere $S^2.$ Clearly, the value of 
\[
H[q]=H_0[q] 
\]
does not depend on the choice of $\Omega ,$ except that, as will be shown
next, by choosing $\Omega >0,$ we can conveniently measure the varying
amount of angular momentum in the fluid. However, unlike the situation for
the standard BVE (which will be discussed next), there is clearly a
difference between the $\Omega >0$ and $\Omega =0$ expressions for the rest
frame kinetic energy of the same generalized barotropic flow.

Dropping the last term which is a constant and rewriting the rest frame
kinetic energy of the fluid (in terms of relative zonal velocity $u_r$ and
meridional $v$ $=v_r),$%
\[
H[q]=\frac 12\int_{S^2}dx\text{ }(u_r^2+v_r^2)+\int_{S^2}dx\text{ }u_ru_p 
\]
we observe that the second term is the projection of relative velocity onto
the velocity of a spherical shell rotating at angular velocity $\Omega ,$
which is proportional to the net angular momentum of the relative flow. This
term cannot be zero for all time since there is no distinguished value for
the spin rate of the solid sphere when angular momentum is exchanged between
the fluid and the sphere. In other words, after choosing some convenient
fixed spin rate $\Omega >0$ to index a rotating frame, the fluid could gain
or loose angular momentum to the sphere and this shows up in the
time-varying inner product 
\[
M_\Omega =\int_{S^2}dx\text{ }u_ru_p. 
\]

To continue with the calculation of energy and momentum in the two
components system, we note that the rest frame kinetic energy of the rigid
sphere is easily calculated to be 
\[
A\Omega _s 
\]
where $A$ is a constant that depends on the radius $R$ and the density of
the sphere and $\Omega _s$ is its changeable angular velocity. Angular
momentum of the sphere in the rest frame is directly related to its kinetic
energy and given by the linear expression 
\[
B\Omega _s. 
\]

By conservation of total kinetic energy (sum of fluid and solid sphere
energy) in the rest frame, the sum 
\[
H[q]+A\Omega _s=const 
\]
even as both terms in the sum changes as $\Omega _s$ changes. Similarly by
conservation of total angular momentum, the sum 
\[
\int_{S^2}dx\text{ }u_ru_p+B\Omega _s=const 
\]
even as both its terms may change over time with $\Omega _s$.

\subsection{The standard BVE}

For pedagogical purposes, we now compare the kinetic energy and angular
momentum expressions for the standard barotropic vorticity model in which
there is neither exchange of energy nor momentum with the sphere (the fluid
component is energetically and torque-wise isolated). This is not the
subject of this paper. The kinetic energy of the fluid in the rest
(inertial) frame is given by 
\[
H_0^{^{\prime }}=\frac 12\int_{S^2}dx\text{ }(u^2+v^2) 
\]
which is superficially the same expression as $H_0$ above but now $%
H_0^{^{\prime }}$ is a constant in time. In a frame that is rotating at
angular velocity $\Omega ,$ the kinetic energy of the fluid is given by 
\[
H^{\prime }[q]=\frac 12\int_{S^2}dx\text{ }(u_r^2+v_r^2)+\int_{S^2}dx\text{ }%
u_ru_p+\frac 12\int_{S^2}dx\text{ }u_p^2 
\]
which is again superficially similar to the expression $H[q]$ but differs
from that earlier expression because it is now a constant. Unlike the
previous situation, the net angular momentum of the fluid relative to this
rotating frame is fixed in time and given by 
\[
M_\Omega ^{\prime }=\int_{S^2}dx\text{ }u_ru_p=const. 
\]

Moreover, there is now a special choice of frame angular velocity $\Omega
^{\prime }$ (or gauge) for which 
\[
M_\Omega ^{\prime }=\int_{S^2}dx\text{ }u_ru_p=0 
\]
for all time, namely the angular velocity $\Omega ^{\prime }$ of a rigidly
rotating spherical shell whose angular momentum equals that of the fluid.
With this choice $\Omega ^{\prime },$ the expression $H^{\prime }[q]$
becomes 
\[
H^{\prime }[q]=\frac 12\int_{S^2}dx\text{ }(u_r^2+v_r^2)+\frac 12\int_{S^2}dx%
\text{ }u_p^2 
\]
which apart from the constant term $\frac 12\int_{S^2}dx$ $u_p^2$ has the
same form as the rest frame kinetic energy $H_0^{\prime }.$ This is the
often stated line that the kinetic energy expression of the standard
barotropic vorticity model (BVE which has constant energy and momentum) has
the same form in any frame, both inertial and rotating ones. Although the
kinetic energies $H^{\prime }[q]=\frac 12\int_{S^2}dx$ $(u_r^2+v_r^2)$ and $%
H_0^{^{\prime }}=\frac 12\int_{S^2}dx$ $(u^2+v^2)$ of the same fluid flow,
have the same form, the relative velocity $(u_r,v_r)$ in the special frame
labeled by $\Omega ^{\prime }$ has zero net angular momentum $M_\Omega
^{\prime }=0$ but the same flow's absolute velocity $(u,v)$ in the inertial
frame has fixed nonzero angular momentum. This fact is properly reflected in
the statistical equilibrium models for the standard BVE by Frederiksen et al 
\cite{Fred} where the conservation of fluid angular momentum is imposed as
an additional microcanonical constraint $M_\Omega ^{\prime }=0.$

\section{Statistical mechanics and enstrophy}

To construct a statistical equilibrium model for this first system, we
should use a formulation that is microcanonical in both the total rest frame
kinetic energy and the total rest frame angular momentum. This formulation
allows kinetic energy and angular momentum to be exchanged between the two
subsystems in the relaxation process towards statistical equilibrium.
However, such a doubly microcanonical statistical ensemble is very
cumbersome to solve.

We therefore assume that the sphere has infinite mass and a fixed angular
velocity $\Omega $, and thus, acts as two related infinite reservoirs of
rest frame kinetic energy $H[q]$ and angular momentum $M_\Omega $ for the
fluid. This simple step is justified in the study of most planetary
atmospheres by the relatively massive planetary spheres in the problem. It
leads to a significant reduction of technical difficulties because the
equilibrium statistical mechanics is now based on a doubly canonical
ensemble in kinetic energy $H[q]$ and angular momentum $M_\Omega $ with
corresponding Lagrange Multipliers or chemical potentials $\beta $ and $%
\alpha .$

We observe that the key expression $H[q]$ in this formulation for
generalized barotropic flows (that exchanges energy and momentum with an
infinite reservoir) is independent of the choice of $\Omega >0,$ precisely
because it is the rest frame kinetic energy of the fluid. Thus, we should
choose $\Omega >0$ to be the fixed angular velocity of the massive solid
sphere. For this choice of $\Omega >0$, the fluctuations of $M_\Omega $
measure the amount of super (resp sub-) rotation in the fluid relative to
the frame in which the solid sphere is fixed. Unlike the standard BVE, there
is really no special choice $\Omega ^{\prime }$ for which the net angular
momentum term $M_{\Omega ^{\prime }}$ vanish for all time.

This problem is still not well-posed because without fixing or limiting the
size of the relative flow in some suitable norm (such as relative enstrophy
in the frame rotating), the energy $H[q]$ and angular momentum $M_\Omega $
can in principle become unbounded at fixed reservoir temperatures. So we
impose the condition of fixed relative enstrophy since it is the square of
the $L_2(S^2)$ norm of the relative vorticity $\omega $. The classical
models based on subjecting enstrophy to a canonical constraint leads to
Gaussian models which are not defined at small absolute values of the
temperature \cite{Kraichnan}.

We note that fixing the relative enstrophy by a microcanonical constraint
means two things: (1) it does not mean that the energy $H[q]$ and angular
momentum $M_\Omega $ are fixed, and (2) it does not mean that the resulting
mixed ensemble is intractable although in general microcanonical constraints
give rise to great analytical difficulties. With one more minor adjustment,
we show in this paper that the resulting models are exactly solvable
spherical Ising and Heisenberg models that can be solved in closed form by
the method of steepest descent.

The last adjustment we make is to couple the two remaining reservoirs into
one, namely fix the statistical temperature $T=\beta ^{-1}$ of a single
energy reservoir instead of having two separate inverse temperatures $\beta $
and $\alpha $ for the energy and angular momentum respectively. This simple
step is justified in the following argument. Expression $H[q]$ for the rest
frame kinetic energy shows that net angular momentum is essentially the
second and independent part of $H[q];$ the first part of $H[q]$ is the
relative kinetic energy in the rotating frame. That is, even as the whole $%
H[q]$ fluctuates in relaxation with respect to the infinite energy
reservoir, the two active parts of $H[q]$, namely the relative kinetic
energy term (1st term) and the angular momentum part (2nd term) exchange
energy constantly even after equilibrium is reached. Thus, the collapse of
two reservoirs into a single energy reservoir in this particular problem,
retains the physically important and statistically independent mechanism of
angular momentum fluctuations.

To summarize this elementary but important material, we note that the
kinetic energy expression used in the derivation of the spin-lattice models
in this paper is just the changeable rest frame kinetic energy of the fluid
(written with respect to a frame rotating at fixed $\Omega >0)$ minus the
constant term $\frac 12\int_{S^2}dx$ $u_p^2,$ 
\[
H[q]=\frac 12\int_{S^2}dx\text{ }(u_r^2+v_r^2)+\int_{S^2}dx\text{ }u_ru_p. 
\]
In general both terms fluctuate independently as the sum $H[q]$ changes in
time due to energy and angular momentum exchanges with the coupled infinite
reservoir. By mapping spins to local vorticity, the first term $\frac
12\int_{S^2}dx$ $(u_r^2+v_r^2)$ by itself gives rise to long range Ising
type spin-lattice models without an external field. Using moreover, the
analogy between magnetic moments and angular momentum, it is easy to see
that the second term $M_\Omega =\int_{S^2}dx$ $u_ru_p$ which represents the
changeable net angular momentum of the fluid (relative to the frame rotating
at fixed $\Omega >0),$ becomes a standard external field term in a
Heisenberg model. It is more convenient to use the Heisenberg models which
are natural vectorial reformulations of the Ising models for generalized
barotropic flow, as shown below.

The principle of angular momentum conservation and the analogy between
angular momentum and magnetic moments are both beautifully illustrated in
the famous Einstein-de Haas experiment where a ferromagnetic rod is
suspended by a thread inside a coil. Upon turning on the current, the rod is
magnetized, that is, its microscopic magnetic spins are aligned. What is
surprising is the fact that the rod rotates inside the coil because the
macroscopic alignment of magnetic moments results in a nonzero net angular
momentum inside the rod due to large numbers of aligned orbiting electrons,
and since no torque is applied to the system, the rod reacts by rotating the
opposite way to conserve angular momentum.

A related process, albeit one that involves phase transitions, is shown in
this paper by obtaining exact solutions for the partition functions of the
spherical Heisenberg models which are the above Heisenberg models plus the
spherical constraint from fixing the relative enstrophy of the flow. At
sufficiently hot negative statistical temperatures (with small absolute
values), we show that the spherical Heisenberg model for generalized
barotropic flows goes through a second order phase transition between
disordered states of local spins (vorticity) at low energy and a global
ordered state at very high energy where the local spins sum to a total
magnetization (net angular momentum) aligned with the rotation axis of the
solid sphere$.$

Reverting to the actual situation where the solid sphere has finite mass,
this phase transition means that the two component fluid -sphere system
undergoes a process very similar to the Einstein-de Haas phenomenon: in the
high energy ordered state the barotropic fluid layer acquires positive net
angular momentum (super-rotates relative to the solid sphere) due to
macroscopic alignment of local vorticity which compares with the orbiting
electrons in the ferromagnetic rod acquiring net angular momentum from
alignment of magnetic moments; the solid sphere slows its rotation rate to
conserve total angular momentum, just as the magnetized rod rotates the
opposite way. Sattinger summarized the phenomenon discussed here succintly
in the phrase ``...many little spins turn into a big spin ..'' \cite{satt}

\section{Heisenberg Model for Barotropic Statistics}

Recall that in the spherical Ising model for barotropic flow, given $N$
fixed mesh points $x_k$ on $S^2$ and the voronoi cells based on this mesh 
\cite{limnebus}, we approximate the relative vorticity by discretizing the
vorticity field as a piecewise constant function, 
\[
\omega (x)=\sum_{j=1}^N{s_jH_j(x)}, 
\]
where $s_j=\omega (x_j)$ and $H_j(x)$ is the characteristic function for the
domain $D_j$, that is \textbf{\ 
\[
H_j(x)=\left\{ 
\begin{array}{rl}
1 & x\in {D_j} \\ 
0 & {otherwise}.
\end{array}
\right. 
\]
}

There is however a more natural vectorial formulation that leads to a
Heisenberg model for barotropic flows on a massive sphere. Instead of
representing the local relative vorticity $\omega (x_j)$ at lattice site $%
x_j $ by a scalar $s_j,$ it is natural to represent it by the vector 
\[
\vec{s}_j=s_j\vec{n}_j 
\]
where $\vec{n}_j$ denotes the outward unit normal to the sphere $S^2$ at $%
x_j.$ Similarly, we represent the spin $\Omega >0$ of the rotating frame by
the vector 
\[
\vec{h}=\frac{2\pi }N\Omega \vec{n} 
\]
where $\vec{n}$ is the outward unit normal at the north pole of $S^2.$
Denoting by $\gamma _{jk}$ the angle subtended at the center of $S^2$ by the
lattice sites $x_j$ and $x_k,$ we obtain the following Heisenberg model for
the total (fixed frame) kinetic energy of a barotropic flow in terms of a
rotating frame at spin rate $\Omega ,$%
\begin{equation}
H_H^N=-\frac 12\sum_{j\neq k}^NJ_{jk}\vec{s}_j\cdot \vec{s}_k+\vec{h}\cdot
\sum_{j=1}^N\vec{s}_j
\end{equation}
where the interaction matrix is now given by the infinite range 
\[
J_{jk}=\frac{16\pi ^2}{N^2}\frac{\ln (1-\cos \gamma _{jk})}{\cos \gamma _{jk}%
}, 
\]
the dot denotes the inner product in $R^3$ and $\vec{h}$ denotes a fixed
external field.

The Kac-Berlin method \cite{Kac} can be modified \cite{Lim06b} to treat the
spherical Heisenberg model which consists of $H_H^N$ and the spherical or
relative enstrophy constraint, 
\[
\frac{4\pi }N\sum_{j=1}^N\vec{s}_j\cdot \vec{s}_j=Q. 
\]
In addition, Stokes theorem implies that it is natural to treat only the
case of zero circulation, 
\[
\frac{4\pi }N\sum_{j=1}^N\vec{s}_j\cdot \vec{n}_j=0. 
\]
Looking ahead, we note the important fact that the following vectorial sum
or magnetization 
\[
\Gamma =\frac{4\pi }N\sum_{j=1}^N\vec{s}_j 
\]
will turn out to be a natural order parameter for the statistics of
barotropic flows on a rotating sphere.

Can the Heisenberg model $H_H^N$ on $S^2$ support phase transitions? More
precisely we check what the Mermin-Wagner theorem has to say about $H_H^N:$
(1) it has spatial dimension $d=2,$ (2) it has a continuous symmetry group,
namely for each element $g\in SO(3)$, 
\[
H_H^N(g\vec{s})=H_H^N(\vec{s}), 
\]
and (3) it has infinite range interaction, that is, for any sequence of
uniform lattices of $N$ sites on $S^2,$ 
\[
\lim_{N\rightarrow \infty }\frac{4\pi }N\sum_{j=1}^NJ_{jk}=-\infty . 
\]
Properties (1) and (2) by themselves would have implied via the
Mermin-Wagner theorem, that $H_H^N$ does not support phase transitions,
since all $d\leq 2,$ finite range models with a continuous symmetry group
cannot have them. However, property (3) violates the finite range condition
of this theorem. Hence, $H_H^N$ on $S^2$ can in principle have phase
transitions in the thermodynamic limit.

It is interesting to compare this Heisenberg model $H_H^N$ on $S^2$ with the
Ising type model $H_N$ for the same barotropic flow in the last section. The
Mermin-Wagner theorem there allows $H_N$ to have phase transitions in the
thermodynamic limit for a different reason: the Ising type interaction $%
J_{jk}=\frac{16\pi ^2}{N^2}\ln (1-\cos \gamma _{jk})$ has finite range
instead of infinite range but $H_N$ does not have a continuous symmetry
group, only the discrete symmetry $Z_2.$

Careful monte-Carlo simulations of this model show that there is one
negative critical temperature in this model, $T_c<0$ for all values of the
spin rate $\Omega $ \cite{dinglim}. An extension of the Kac-Berlin method to
the spherical Heisenberg model for Barotropic flows on a rotating sphere
will show that BEC transitions through a symmetry-breaking Goldstone mode to
the single ground state $\psi _{10},$ occurs for sufficiently high kinetic
energies or small negative values of the temperature $T$.

\section{Solution of the spherical Heisenberg model for $\Omega >0$}

The family of Heisenberg models $H_H^N$ derived above for the barotropic
fluid - solid sphere system in a frame rotating at angular velocity $\Omega
>0$ have external fields $\vec{h}(N)=\frac{2\pi }N\Omega \vec{n}$ and
infinite range interactions 
\[
J_{jk}=\frac{16\pi ^2}{N^2}\frac{\ln (1-\cos \gamma _{jk})}{\cos \gamma _{jk}%
}. 
\]
Combining it with the vectorial spherical constraint, 
\[
\frac{4\pi }N\sum_{j=1}^N\vec{s}_j\cdot \vec{s}_j=Q, 
\]
we obtain an extension of Kac's spherical model \cite{Kac} to a
one-parameter family of spherical Heisenberg models which is parametrized by
the size $N$ of the Voronoi lattice on $S^2.$

This family of spherical Heisenberg models for barotropic vortex statistics
allows us to model the thermal interactions between local relative vorticity 
$\omega (x)$ and a kinetic energy reservoir at any fixed temperature $T.$
The spherical constraint enforces the microcanonically fixed relative
enstrophy $Q>0$ but allows angular momentum in each of the three principal
directions to change. Similar to the equilibrium condensation process found
in the case $\Omega =0$ for the spherical Ising model [Lim06a], kinetic
energy of barotropic flow settles into a Goldstone symmetry-breaking ground
state at very small negative temperatures $T_c<T<0$ (associated with
extremely large energies). Unlike the $\Omega =0,$ there is no 3-fold
degeneracy in the Goldstone modes and only the mode $\psi _{10}$ which
carries angular momentum that is aligned with the rotation axis $\Omega \vec{%
n},$ has a large amplitude.

The exact solution of the spherical Heisenberg models $H_H^N$ proceeds along
similar lines to the Kac-Berlin method for the spherical Ising model. In the
thermodynamic or continuum limit as $N\rightarrow \infty ,$ the partition
function is calculated using Laplace's integral form, 
\begin{eqnarray*}
Z_H^N &\propto &\int D(\vec{s})\exp \left( -\beta H_H^N(\vec{s})\right)
\delta \left( Q\frac N{4\pi }-\sum_{j=1}^N\vec{s}_j\cdot \vec{s}_j\right) \\
&=&\int D(\vec{s})\exp \left( -\beta H_H^N(\vec{s})\right) \left( \frac
1{2\pi i}\int_{a-i\infty }^{a+i\infty }d\eta \exp \left( \eta \left( Q\frac
N{4\pi }-\sum_{j=1}^N\vec{s}_j\cdot \vec{s}_j\right) \right) \right) \\
&=&\int D(\vec{s})\exp \left( \frac \beta 2\sum_{j\neq k}^NJ_{jk}\vec{s}%
_j\cdot \vec{s}_k-\beta \vec{h}\cdot \sum_{j=1}^N\vec{s}_j\right) \left(
\frac 1{2\pi i}\int_{a-i\infty }^{a+i\infty }d\eta \exp \left( \eta \left( N-%
\frac{4\pi }Q\sum_{j=1}^N\vec{s}_j\cdot \vec{s}_j\right) \right) \right)
\end{eqnarray*}
where moreover, the microstate 
\[
\vec{s}=\{\vec{s}_1,...,\vec{s}_N\}\in R^{3N} 
\]
satisfies the zero circulation condition 
\[
\sum_{j=1}^N\vec{s}_j\cdot \vec{n}_j=0. 
\]

Thus, 
\begin{eqnarray*}
Z_H^N &\propto &\left( \int D(\vec{s})\exp \left( \frac \beta 2\sum_{j\neq
k}^NJ_{jk}\vec{s}_j\cdot \vec{s}_k-\beta \vec{h}\cdot \sum_{j=1}^N\vec{s}%
_j\right) \int_{a-i\infty }^{a+i\infty }\frac{d\eta }{2\pi i}\exp \left(
\eta \left( N-\frac{4\pi }Q\sum_{j=1}^N\vec{s}_j\cdot \vec{s}_j\right)
\right) \right) \\
&=&\int D(\vec{s})\int_{a-i\infty }^{a+i\infty }\frac{d\eta }{2\pi i}\exp
\left( N\left( 
\begin{array}{c}
\eta -\frac{4\pi }{QN}\eta \sum_{j=1}^N\vec{s}_j\cdot \vec{s}_j \\ 
+\frac \beta {2N}\sum_{j\neq k}^NJ_{jk}\vec{s}_j\cdot \vec{s}_k-\frac \beta N%
\vec{h}\cdot \sum_{j=1}^N\vec{s}_j
\end{array}
\right) \right) \\
&=&\int D(\vec{s})\int_{a-i\infty }^{a+i\infty }\frac{d\eta }{2\pi i}\exp
\left( N\left( \eta -\frac 1N\sum_{j\neq k}^NK_{jk}(Q,\beta ,\eta )\text{ }%
\vec{s}_j\cdot \vec{s}_k-\frac \beta N\vec{h}\cdot \sum_{j=1}^N\vec{s}%
_j\right) \right)
\end{eqnarray*}
where 
\[
K_{jk}(Q,\beta ,\eta )=\left\{ 
\begin{array}{cc}
\frac{4\pi }Q\eta & j=k \\ 
-\frac \beta 2J_{jk} & j\neq k
\end{array}
\right\} . 
\]

To evaluate the Gaussian integrals in $Z_H^N,$ we expand the relative
vorticity vectorfield again, this time, in terms of the spherical harmonics, 
\[
\vec{\omega}(x)=\sum_{l=1,m=-l}^{\infty ,l}\alpha _{lm}\psi _{lm}(x)\vec{n}%
(x) 
\]
where $\vec{n}(x)$ is the outward unit normal to $S^2$ at $x.$ We stress
that this expansion need not include $\psi _{00}(x)=c$ because of the zero
circulation condition on microstates $\vec{s}.$

Solution of the Gaussian integrals requires diagonalizing the interaction in 
$H_H^N$ in terms of the spherical harmonics $\left\{ \psi _{lm}\right\}
_{l=1,}^\infty $ which are natural Fourier modes for Laplacian eigenvalue
problems on $S^2$ with zero circulation: 
\begin{eqnarray*}
\vec{s}_j &=&\vec{n}_j\sum_{l=1}^\infty \sum_{m=-l}^l\alpha _{lm}\psi
_{lm}(x_j) \\
-\frac 12\sum_{j\neq k}^N &&J_{jk}\text{ }\vec{s}_j\cdot \vec{s}_k=\frac
12\sum_{l=1}^\infty \sum_{m=-l}^l\lambda _{lm}\alpha _{lm}^2 \\
\vec{h}\cdot \sum_{j=1}^N &&\vec{s}_j=\frac 12\Omega C\alpha _{10}
\end{eqnarray*}
where the eigenvalues of the Green's function for the Laplace-Beltrami
operator on $S^2$ are 
\[
\lambda _{lm}=\frac 1{l(l+1)},\text{ }l=1,...,\sqrt{N},\text{ }%
m=-l,...,0,...,l 
\]
and $\alpha _{lm}$ are the corresponding amplitudes. Thus, 
\[
\frac 1N\sum_{j\neq k}^NK_{jk}(Q,\beta ,\eta )\text{ }\vec{s}_j\cdot \vec{s}%
_k=\sum_{l=1}^\infty \sum_{m=-l}^l\left( \frac \beta {2N}\lambda _{lm}+\frac
\eta Q\right) \alpha _{lm}^2 
\]
and

\begin{eqnarray*}
Z_H^N &\propto &\int D(\vec{s})\int_{a-i\infty }^{a+i\infty }\frac{d\eta }{%
2\pi i}\exp \left( N\left( \eta -\frac 1N\sum_{j\neq k}^NK_{jk}(Q,\beta
,\eta )\text{ }\vec{s}_j\cdot \vec{s}_k-\frac \beta N\vec{h}\cdot
\sum_{j=1}^N\vec{s}_j\right) \right) \\
&=&\int \prod_{m=-1}^1d\alpha _{1m}\int D_{l\geq 2}(\alpha )\int_{a-i\infty
}^{a+i\infty }\frac{d\eta }{2\pi i}\exp \left\{ N\left[ 
\begin{array}{c}
\eta -\sum_{l=2}^\infty \sum_{m=-l}^l\left( \frac \beta {2N}\lambda
_{lm}+\frac \eta Q\right) \alpha _{lm}^2 \\ 
-\left( \frac \beta {4N}+\frac \eta Q\right) \sum_{m=-1}^1\alpha
_{1m}^2-\frac \beta {2N}\Omega C\alpha _{10}
\end{array}
\right] \right\}
\end{eqnarray*}
\[
=\int \prod_{m=-1}^1d\alpha _{1m}\int_{a-i\infty }^{a+i\infty }\frac{d\eta }{%
2\pi i}\exp \left\{ N\left[ \eta -\left( \frac \beta {4N}+\frac \eta
Q\right) \sum_{m=-1}^1\alpha _{1m}^2-\frac \beta {2N}\Omega C\alpha
_{10}\right] \right\} 
\]
\[
\int D_{l\geq 2}(\alpha )\exp \left( -N\sum_{l=2}^\infty \sum_{m=-l}^l\left(
\frac \beta {2N}\lambda _{lm}+\frac \eta Q\right) \alpha _{lm}^2\right) 
\]
where the order of integration of the term $\exp \left( -N\sum_{l=2}^\infty
\sum_{m=-l}^l\frac \eta Q\alpha _{lm}^2\right) $ can be interchanged by
choosing $\func{Re}(\eta )=a>0$ large enough.

\subsection{ Restricted partition function and non-ergodic modes}

Next we write the problem in terms of the restricted partition function $%
Z_H^N(\alpha _{10},\alpha _{1,\pm 1};\beta ,Q,\Omega ),$ that is, 
\begin{eqnarray*}
Z_H^N(\beta ,Q,\Omega ) &\propto &\int \prod_{m=-1}^1d\alpha _{1m}\text{ }%
Z_H^N(\alpha _{10},\alpha _{1,\pm 1};\beta ,Q,\Omega ) \\
&=&\int \prod_{m=-1}^1d\alpha _{1m}\int_{a-i\infty }^{a+i\infty }\frac{d\eta 
}{2\pi i}\text{ }\exp \left\{ N\left[ \eta -\left( \frac \beta {4N}+\frac
\eta Q\right) \sum_{m=-1}^1\alpha _{1m}^2-\frac \beta {2N}\Omega C\alpha
_{10}\right] \right\}
\end{eqnarray*}
\[
\int D_{l\geq 2}(\alpha )\exp \left( -N\sum_{l=2}^\infty \sum_{m=-l}^l\left(
\frac \beta {2N}\lambda _{lm}+\frac \eta Q\right) \alpha _{lm}^2\right) . 
\]

Because of non-ergodicity of the condensed modes, we should not integrate
over the ordered modes in this problem, namely $\alpha _{1m}$, which are the
amplitudes of the 3-fold degenerate ground modes $\psi _{1m}$ that carry
global angular momentum. This often used physical argument in the condensed
matter literature will for the first time be turned into a rigorous proof
here. A pertinent and important question arises at this point \cite{xuer}:
how many and what are the condensed modes in any given spherical model? We
will show later that only one single class of modes can have nonzero
amplitudes in the condensed phase of this problem, namely those belonging to
the meridional wave number $l=1.$

The statistics of the problem are therefore completely determined by the
restricted partition function $Z_H^N(\alpha _{10},\alpha _{1,\pm 1};\beta
,Q,\Omega ).$ Amplitudes $\alpha _{10},\alpha _{1,\pm 1}$ of the ordered
modes appear as parameters in this restricted partition function, and will
have to be evaluated separately.

Standard Gaussian integration is used to evaluate the last integral, which
yields, after scaling $\beta ^{\prime }N=\beta ,$ 
\[
\int_{l\geq 2}D(\alpha )\exp \left( -\sum_{l=2}\sum_{m=-l}^l\left( \frac{%
\beta ^{\prime }N\lambda _{lm}}2+\frac{N\eta }Q\right) \alpha _{lm}^2\right)
=\prod_{l=2}^{\sqrt{N}}\prod_{m=-l}^l\left( \frac \pi {\frac{N\eta }Q+\frac{%
\beta ^{\prime }N}2\lambda _{lm}}\right) ^{1/2}, 
\]
provided the physically significant Gaussian conditions hold: for $l\geq 2,$ 
\begin{equation}
\frac{\beta ^{\prime }\lambda _{lm}}2+\frac \eta Q=\frac{\beta ^{\prime }}{%
2l(l+1)}+\frac \eta Q>0.  \label{gauss2}
\end{equation}
Then the partition function takes the form 
\[
Z_H^N(\alpha _{10},\alpha _{1,\pm 1};\beta ,Q,\Omega )\propto \frac 1{2\pi
i}\int_{a-i\infty }^{a+i\infty }d\eta \text{ }\exp \left\{ N\left[ 
\begin{array}{c}
\eta -\left( \frac{\beta ^{\prime }}4+\frac \eta Q\right)
\sum_{m=-1}^1\alpha _{1m}^2-\frac{\beta ^{\prime }}2\Omega C\alpha _{10} \\ 
-\frac 1{2N}\sum_{l=2}\sum_m\ln \left( \frac{N\eta }Q+\frac{\beta ^{\prime }N%
}2\lambda _{lm}\right)
\end{array}
\right] \right\} 
\]
where the free energy per site evaluated at the most probable macrostate is $%
-\frac 1{\beta ^{\prime }}$ $F(\eta (\beta ^{\prime }),Q,\beta ^{\prime })\,$
with 
\[
F(\eta (\beta ^{\prime }),Q,\beta ^{\prime })=\eta (\beta ^{\prime })\left[
1-\frac 1Q\sum_{m=-1}^1\alpha _{1m}^2\right] -\frac{\beta ^{\prime }}%
4\sum_{m=-1}^1\alpha _{1m}^2-\frac{\beta ^{\prime }}2\Omega C\alpha _{10} 
\]
\[
-\frac 1{2N}\sum_{l=2}\sum_m\ln \left( \frac{N\eta }Q+\frac{\beta ^{\prime }N%
}2\lambda _{lm}\right) . 
\]

\subsection{Planck's theorem, Saddle points and the Themodynamic limit}

Provided that the saddle point $\eta (\beta ^{\prime })$ can be determined
at given inverse temperature $\beta ^{\prime },$ Planck's theorem states
that the thermodynamically stable (most probable) macrostate is given by the
maximum of the expression $F(\eta (\beta ^{\prime }),Q,\beta ^{\prime }).$
At positive temperatures, the structure of this expression where it concerns
the ground modes $\alpha _{1m},$ namely, 
\[
\chi (\alpha _{10},\alpha _{1,\pm 1};\beta ,Q,\Omega )=\eta (\beta ^{\prime
})\left[ 1-\frac 1Q\sum_{m=-1}^1\alpha _{1m}^2\right] -\left[ \frac{\beta
^{\prime }}4\sum_{m=-1}^1\alpha _{1m}^2+\frac{\beta ^{\prime }}2\Omega
C\alpha _{10}\right] , 
\]
and the fact that the saddle point $\eta (\beta ^{\prime })$ must be
positive, suggests that for any positive value of the saddle point, the
expression $\chi $ and therefore $F(\eta (\beta ^{\prime }),Q,\beta ^{\prime
})$ is maximized by $\sum_{m=-1}^1\alpha _{1m}^2=0$ for all $\beta ^{\prime
}>0$ when planetary spin $\Omega $ is small, and by $\alpha _{10}<0$ for
large $\beta ^{\prime }>0$ when planetary spin $\Omega $ is large. At
negative temperatures, we expect to find a finite critical point where the
two opposing parts of $\chi $ are balanced. In order to prove that these
heuristic expectations are valid, we will solve the restricted partition
function in closed form by the method of steepest descent.

The saddle point condition gives one equation for the determination of four
variables $\eta ,\alpha _{1m}$ in terms of inverse temperature $\beta
^{\prime }$ and relative enstrophy $Q,$%
\begin{equation}
0=\frac{\partial F}{\partial \eta }=\left( 1-\frac 1Q\sum_{m=-1}^1\alpha
_{1m}^2\right) -\frac 1{2NQ}\sum_{l=2}\sum_m\left( \frac{\eta (\beta
^{\prime })}Q+\frac{\beta ^{\prime }}2\lambda _{lm}\right) ^{-1}
\label{sadd3}
\end{equation}
where $\eta =\eta (\beta ^{\prime })$ is taken to be the value of the saddle
point. Note that it does not depend on the planetary spin rate $\Omega >0.$
We note in passing that the same equation holds in the $\Omega =0$ case.
There are two natural subcases for the saddle point condition, namely, (A)
the disordered phase (for $|T^{\prime }|\gg 1)$ where equation (\ref{sadd3})
has finite solution $\eta (\beta ^{\prime })>0,$ and $\alpha _{1m}=0$ for $%
m=-1,0,1;$ and (B) the ordered or condensed phase (for $|T^{\prime }|\ll 1)$
where equation (\ref{sadd3}) has finite solution $\eta (\beta ^{\prime })>0$
only when $\alpha _{1m}\neq 0$ for some $m.$ In case (A) which will be
solved below, there is no need to invoke additional equations of state as
the amplitudes $\alpha _{1m}=0$ for $m=-1,0,1.$

Case (B) requires three more conditions to determine the three amplitudes $%
\alpha _{1m}$ and the saddle point $\eta (\beta ^{\prime })>0.$ They are
provided by equations of state (or Planck's theorem) for the condensed phase
(which do not hold in the disordered phase): 
\begin{eqnarray}
0 &=&\frac{\partial F}{\partial \alpha _{10}}=-\left( \frac{2\eta (\beta
^{\prime })}Q+\frac{\beta ^{\prime }}2\right) \alpha _{10}-\frac{\beta
^{\prime }}2\Omega C  \label{state1} \\
0 &=&\frac{\partial F}{\partial \alpha _{1,\pm 1}}=-\left( \frac{2\eta
(\beta ^{\prime })}Q+\frac{\beta ^{\prime }}2\right) \alpha _{1,\pm 1}.
\label{state23}
\end{eqnarray}
Thus, a coupled system of four algebraic equations (\ref{sadd3}), (\ref
{state1}), (\ref{state23}) determines four unknowns in terms of the
planetary spin $\Omega >0,$ the relative enstrophy $Q>0$ and the scaled
inverse temperature $\beta ^{\prime }.$ The last two equations of state for $%
\alpha _{1,\pm 1}$ implies that either 
\[
\alpha _{1,\pm 1}=0\text{ or }\left( \frac{2\eta (\beta ^{\prime })}Q+\frac{%
\beta ^{\prime }}2\right) =0. 
\]
The first equation of state differs from the other two; this represents
reduction of the $SO(3)$ symmetry that existed in the $\Omega =0$ case to $%
S^1$ symmetry in the case of nonzero planetary spin. Together these three
equations of state imply that when $\Omega >0,$ the only possible solution
is without tilt, 
\begin{eqnarray}
\alpha _{10} &=&-\frac{\beta ^{\prime }\Omega C}2\left( \frac{2\eta (\beta
^{\prime })}Q+\frac{\beta ^{\prime }}2\right) ^{-1}\neq 0,  \label{order} \\
\alpha _{1,\pm 1} &=&0.  \nonumber
\end{eqnarray}
These values of $\alpha _{lm}$ will be substituted back into the saddle
point condition (\ref{sadd3}) to yield a single equation that will be solved
below.

The Gaussian conditions (\ref{gauss2}) imply that for $l>1,$ 
\[
\frac{\beta ^{\prime }}{2l(l+1)}+\frac{\eta (\beta ^{\prime })}Q>0. 
\]

The critical temperature can be obtained from the saddle point condition:
(A) in the disordered phase at large $|T|$, 
\begin{equation}
1=\lim_{N\rightarrow \infty }\frac 1{2NQ}\sum_{l=2}^{\sqrt{N}%
}\sum_{m=-l}^l\left( \frac{\eta (\beta ^{\prime })}Q+\frac{\beta ^{\prime }}{%
2l(l+1)}\right) ^{-1}  \label{casea}
\end{equation}
where the large $N$ limit on the RHS is well-defined and finite for any
finite $|\beta ^{\prime }|$ provided 
\begin{equation}
\eta (\beta ^{\prime })\geq \eta ^{*}=\frac{|\beta ^{\prime }|Q}4>0,
\label{con1}
\end{equation}
because then, each term $(l\geq 2)$ in the sum is majorized: for negative
temperatures, 
\[
\left( \frac{\eta (\beta ^{\prime })}Q+\frac{\beta ^{\prime }}{2l(l+1)}%
\right) ^{-1}\leq \left( -\frac{\beta ^{\prime }}4+\frac{\beta ^{\prime }}{%
2l(l+1)}\right) ^{-1}\leq \left( -\frac{\beta ^{\prime }}6\right) ^{-1}, 
\]
and for positive temperatures, 
\[
\left( \frac{\eta (\beta ^{\prime })}Q+\frac{\beta ^{\prime }}{2l(l+1)}%
\right) ^{-1}\leq \left( \frac{\beta ^{\prime }}4-\frac{\beta ^{\prime }}{%
2l(l+1)}\right) ^{-1}\leq \left( \frac{\beta ^{\prime }}6\right) ^{-1}; 
\]
and the corresponding expressions have well-defined positive limits, i.e.,
for all negative and finite $\beta ^{\prime },$ 
\begin{equation}
\lim_{N\rightarrow \infty }\frac 1{2NQ}\sum_{l=2}^{\sqrt{N}%
}\sum_{m=-l}^l\left( -\frac{\beta ^{\prime }}4+\frac{\beta ^{\prime }}{%
2l(l+1)}\right) ^{-1}<\infty ,  \label{thermo}
\end{equation}
and for all positive and finite $\beta ^{\prime },$%
\[
\lim_{N\rightarrow \infty }\frac 1{2NQ}\sum_{l=2}^{\sqrt{N}%
}\sum_{m=-l}^l\left( \frac{\beta ^{\prime }}4-\frac{\beta ^{\prime }}{2l(l+1)%
}\right) ^{-1}<\infty . 
\]

And (B) in the ordered phase at small $|T|,$%
\begin{equation}
\left( 1-\frac 1Q\alpha _{10}^2\right) =\lim_{N\rightarrow \infty }\frac
1{2NQ}\sum_{l=2}\sum_m\left( \frac{\eta (\beta ^{\prime })}Q+\frac{\beta
^{\prime }}{2l(l+1)}\right) ^{-1}  \label{caseb}
\end{equation}
where a similar argument proves that the RHS is well-defined and finite
provided $\eta (\beta ^{\prime })\geq \eta ^{*}.$

This proves that the thermodynamic or continuum limit of the spherical
Heisenberg model $H_H^N$ is well-defined for all negative temperatures
because it turns out (and is shown below) that the saddle point satisfies (%
\ref{con1}) for the disordered as well as the ordered phases. Later we will
show that this thermodynamic limit exists for all positive temperatures as
well.

The large $|T|$ or small $|\beta ^{\prime }|$ saddle point condition in case
(A), 
\begin{equation}
\lim_{N\rightarrow \infty }\frac 1{2N}\sum_{l=2}^{\sqrt{N}%
}\sum_{m=-l}^l\left( \frac{\eta (\beta ^{\prime })}Q+\frac{\beta ^{\prime }}{%
2l(l+1)}\right) ^{-1}=Q,  \label{sadsad}
\end{equation}
can be solved and has the property that $\eta (\beta ^{\prime })\searrow 1$
as $|\beta ^{\prime }|\rightarrow 0.$

In case (B), when $|\beta ^{\prime }|$ is large, we discuss (i) $\beta
^{\prime }<0$ and (ii) $\beta ^{\prime }>0$ separately.

\subsection{Negative critical temperature}

For case (i) $\beta ^{\prime }<0$, a point is reached at $\beta _c^{\prime }$
where 
\[
-\infty <\beta _c^{\prime }(Q)=\lim_{N\rightarrow \infty }\frac
1{QN}\sum_{l=2}^{\sqrt{N}}\sum_{m=-l}^l\left( \lambda _{lm}-\frac 12\right)
^{-1}<0, 
\]
such that for $\beta ^{\prime }<\beta _c^{\prime }(Q)<0,$%
\[
\lim_{N\rightarrow \infty }\frac 1{\beta ^{\prime }NQ}\sum_{l=2}\sum_m\left(
-\frac 12+\lambda _{lm}\right) ^{-1}<1. 
\]
(We note the significant fact that $T_c^{\prime }(Q)$ depends linearly on
the relative enstrophy $Q$ but does not depend on $\Omega .)$ In other
words, the extreme saddle point 
\[
\eta ^{*}=-\frac{\beta ^{\prime }Q}4 
\]
is no longer adequate to solve (\ref{casea}) for $\beta ^{\prime }<\beta
_c^{\prime }(Q)<0;$ any larger value $\eta >\eta ^{*}$ does not work either
because then 
\[
\lim_{N\rightarrow \infty }\frac 1{2N}\sum_{l=2}^{\sqrt{N}%
}\sum_{m=-l}^l\left( \frac \eta Q+\frac{\beta ^{\prime }}{2l(l+1)}\right)
^{-1}<Q 
\]
for $\beta ^{\prime }<\beta _c^{\prime }(Q)<0.$ It remains to check that $%
\eta <\eta ^{*}$ cannot be used. This is due to the fact that 
\[
\lim_{N\rightarrow \infty }\frac 1{2N}\sum_{l=2}^{\sqrt{N}%
}\sum_{m=-l}^l\left( \frac{\eta ^{*}}Q+\frac{\beta _c^{\prime }}{2l(l+1)}%
\right) ^{-1}=Q 
\]
which implies 
\[
\lim_{N\rightarrow \infty }\frac 1{2N}\sum_{l=2}^{\sqrt{N}%
}\sum_{m=-l}^l\left( \frac \eta Q+\frac{\beta ^{\prime }}{2l(l+1)}\right)
^{-1}>Q 
\]
if $\eta <\eta ^{*}$ and $\beta ^{\prime }<\beta _c^{\prime }<0.$

From this discussion of (i) $\beta ^{\prime }\leq \beta _c^{\prime }<0,$ and
after substituting the nonzero solution (\ref{order}) of the equations of
state back into the saddle point equation, 
\begin{equation}
\left( 1-\frac 1Q\alpha _{10}^2\right) =\lim_{N\rightarrow \infty }\frac
1{2NQ}\sum_{l=2}\sum_m\left( \frac{\eta (\beta ^{\prime })}Q+\frac{\beta
^{\prime }}{2l(l+1)}\right) ^{-1},  \label{sadord}
\end{equation}
we derive a single equation, 
\begin{equation}
\left( 1-\frac{(\beta ^{\prime })^2\Omega ^2C^2}{16Q}\left( \frac{\eta
(\beta ^{\prime },\Omega ,Q)}Q+\frac{\beta ^{\prime }}4\right) ^{-2}\right)
=\lim_{N\rightarrow \infty }\frac 1{2NQ}\sum_{l=2}\sum_m\left( \frac{\eta
(\beta ^{\prime },\Omega ,Q)}Q+\frac{\beta ^{\prime }}{2l(l+1)}\right) ^{-1}
\label{sadalpha}
\end{equation}
for the saddle point $\eta (\beta ^{\prime },\Omega ,Q)\geq \eta ^{*}$ when $%
\beta ^{\prime }\leq \beta _c^{\prime }<0.$

The RHS of this equation can be made larger (resp. smaller) than $1$ by
choosing $\eta (\beta ^{\prime })<\eta ^{*}$ $(>\eta ^{*}$ resp.) and since $%
Q=\sum_{l=1}\sum_m\alpha _{lm}^2$ is the relative enstrophy, we must have 
\[
0\leq \left( 1-\frac 1Q\alpha _{10}^2\right) \leq 1 
\]
which means that its LHS lie between $0$ and $1.$ Thus, by choosing a
suitable $\eta (\beta ^{\prime })\geq \eta ^{*}$ we should be able to
satisfy (\ref{sadalpha}) for $\beta ^{\prime }\leq \beta _c^{\prime }<0$. It
remains to check that this is consistent with the property $\alpha
_{10}^2\leq Q$ of the ordered solution (\ref{order}), that is, for all $%
\Omega ,$ 
\begin{equation}
\frac{(\beta ^{\prime })^2\Omega ^2C^2}{16Q}\left( \frac{\eta (\beta
^{\prime },\Omega ,Q)}Q+\frac{\beta ^{\prime }}4\right) ^{-2}\leq 1.
\label{alpha}
\end{equation}

Thus, to prove that the saddle point condition (\ref{sadalpha}) has
solutions $\eta (\beta ^{\prime },\Omega ,Q)\geq \eta ^{*}$ for all $\Omega
>0$, $Q>0,$ and for all $\beta ^{\prime }\leq \beta _c^{\prime }(Q)<0,$ it
is sufficient to note that for fixed $\beta ^{\prime }\leq \beta _c^{\prime
}(Q),$ its RHS($\eta (\beta ^{\prime }))<1,$ decreases as $\eta $ $>\eta
^{*} $ increases, while its LHS($\eta (\beta ^{\prime }))$ $<1,$ increases
as $\eta $ $>\eta ^{*}$ increases; and in such a way that RHS($\eta (\beta
^{\prime }))$ is surjective on the interval $(0,1)$ with $\lim_{\eta (\beta
^{\prime })\nearrow \infty }RHS(\eta (\beta ^{\prime }))=0$ for any fixed $%
\beta ^{\prime }<\beta _c^{^{\prime }},$ and $RHS(\eta ^{*}(\beta _c^{\prime
}))=1,$ and LHS($\eta (\beta ^{\prime }))$ is surjective on $(0,1)$ with $%
\lim_{\eta (\beta ^{\prime })\nearrow \infty }LHS(\eta (\beta ^{\prime }))=1$
for any fixed $\beta ^{\prime }<\beta _c^{^{\prime }},$and $LHS(\bar{\eta}%
(\beta ^{\prime }))=0$ for the solution 
\[
\bar{\eta}(\beta ^{\prime })=-\frac{\beta ^{\prime }\Omega C\sqrt{Q}}4-\frac{%
\beta ^{\prime }Q}4>\eta ^{*} 
\]
of 
\[
\frac{(\beta ^{\prime })^2\Omega ^2C^2}{16Q}\left( \frac{\bar{\eta}(\beta
^{\prime },\Omega ,Q)}Q+\frac{\beta ^{\prime }}4\right) ^{-2}=1. 
\]
Condition (\ref{alpha}) then implies that for all $\Omega >0$, $Q>0,$ and
for all $\beta ^{\prime }\leq \beta _c^{\prime }(Q)<0,$the saddle point $%
\eta (\beta ^{\prime },\Omega ,Q)$ satisfies 
\[
\eta (\beta ^{\prime },\Omega ,Q)\geq -\frac{\beta ^{\prime }\Omega C\sqrt{Q}%
}4-\frac{\beta ^{\prime }Q}4>\eta ^{*} 
\]
which proves the reflection property in the following remark.

\textbf{Remark 1:} \textsl{Since the extreme saddle point } 
\[
\eta ^{*}=-\frac{\beta ^{\prime }Q}4 
\]
\textsl{satisfies the saddle point conditions (\ref{casea}) and (\ref{caseb}%
) only at the single value of the temperature }$T_c^{^{\prime }}<0$ \textsl{%
that separates the disordered phase from the condensed phase, but not at
other }$T<0,$\textsl{\ we have shown that the usual phenomenon known as, 
\textit{sticking of the saddle point in the ordered phase,} does not hold
here. A more appropriate label for this new saddle point behaviour seen in
the spherical Heisenberg models for barotropic flows on a rotating sphere,
is \textit{jumping and reflection of the saddle point at the negative
critical point. }Indeed the proof above shows that,}\textit{\ for all }$%
\Omega >0$\textit{\ and }$Q>0,$\textit{\ and for all }$\beta ^{\prime
}<\beta _c^{\prime }(Q)<0$\textit{, the saddle point }$\eta (\beta ^{\prime
})\geq -\frac{\beta ^{\prime }\Omega C\sqrt{Q}}4-\frac{\beta ^{\prime }Q}%
4>\eta ^{*}.$

We summarize the above results in the physical theorem:

\textbf{Theorem 1:} (A) \textit{For all }$\Omega >0$\textit{\ and }$Q>0,$%
\textit{\ the quantity } 
\[
\beta _c^{\prime }(Q,N)=\frac 1{QN}\sum_{l=2}^{\sqrt{N}}\sum_{m=-l}^l\left(
\lambda _{lm}-\frac 12\right) ^{-1}<0 
\]
\textit{has a well-defined and finite limit, called the critical inverse
temperature,} 
\[
\beta _c^{\prime }(Q)=\lim_{N\rightarrow \infty }\beta _c^{\prime
}(Q,N)>-\infty , 
\]
\textit{that is independent of the rate of spin }$\Omega .$\textit{\ }

\textit{(B)} \textit{Moreover, the thermodynamic limit exists for the
spherical Heisenberg models }$H_H^N$ \textit{in the sense that for any }$Q>0$%
\textit{\ and }$\Omega >0,$\textit{\ the saddle point conditions,} 
\begin{eqnarray*}
1 &=&\lim_{N\rightarrow \infty }\frac 1{2NQ}\sum_{l=2}\sum_m\left( \frac{%
\eta (\beta ^{\prime })}Q+\frac{\beta ^{\prime }}{2l(l+1)}\right) ^{-1}\text{
} \\
&&\left( 1-\frac 1Q\alpha _{10}^2\right) =\lim_{N\rightarrow \infty }\frac
1{2NQ}\sum_{l=2}\sum_m\left( \frac{\eta (\beta ^{\prime })}Q+\frac{\beta
^{\prime }}{2l(l+1)}\right) ^{-1},
\end{eqnarray*}
\textit{are well-defined and finite, and the saddle point satisfies the
condition } 
\[
\eta (\beta ^{\prime })\geq \eta ^{*}=-\frac{\beta ^{\prime }Q}4>0 
\]
\textit{\ for all }$\beta ^{\prime }<0.$

\textit{(C) For all }$\Omega >0$\textit{\ and }$Q>0,$\textit{\ and for all }$%
\beta ^{\prime }<\beta _c^{\prime }(Q)<0$\textit{, the ordered phase takes
the form of the tiltless (}$\alpha _{1,\pm 1}=0)$\textit{\ ground mode }$%
\alpha _{10}(\beta ^{\prime },\Omega ,Q)\psi _{10}$ \textit{with amplitude} 
\[
\alpha _{10}=-\frac{\beta ^{\prime }\Omega C}2\left( \frac{2\eta (\beta
^{\prime })}Q+\frac{\beta ^{\prime }}2\right) ^{-1}>0, 
\]
\textit{which implies that it is aligned with the rotation }$\Omega >0$ 
\textit{(super-rotating)} \textit{and is linear in }$\Omega .$

\subsection{Positive temperature}

For case (ii) $\beta ^{\prime }>0$, we note that the Gaussian conditions (%
\ref{gauss2}) are automatically satisfied since $\eta (\beta ^{\prime },Q)/Q$
$>0$ is required of the saddle point of the equation

\begin{equation}
\left( 1-\frac 1Q\alpha _{10}^2\right) =\lim_{N\rightarrow \infty }\frac
1{2NQ}\sum_{l=2}\sum_m\left( \frac{\eta (\beta ^{\prime })}Q+\frac{\beta
^{\prime }}{2l(l+1)}\right) ^{-1},  \label{saddd}
\end{equation}
We deduce from the saddle point condition (\ref{saddd}), that for any $Q>0,$
and any positive $\beta ^{\prime }(Q)<\infty ,$ there is a saddle point $%
\eta (\beta ^{\prime },Q)>0$ associated with the disordered phase $\alpha
_{10}=0.$ Otherwise, there is a finite critical point $\beta _{cc}^{\prime
}(Q)>0$ that satisfies the equation, 
\[
\beta _{cc}^{\prime }=\lim_{N\rightarrow \infty }\frac 1{2NQ}\sum_{l=2}^{%
\sqrt{N}}\sum_{m=-l}^l2l(l+1)<\infty , 
\]
which is a contradiction since the sum on the RHS does not converge. The
extreme case, namely $\eta (\beta ^{\prime })=0,$ for the saddle point
condition holds at precisely one point, that is, for $\beta ^{\prime
}=\infty .$

We will show next that (\ref{saddd}) has more than one saddle points at all
positive temperatures. In addition to the disordered phase solution found
above, the pure ground mode phase is a saddle point $\eta ^{\prime },\alpha
_{10}<0$ of (\ref{saddd}). Using the solution (\ref{order}) of the equation
of state for amplitude $\alpha _{10}$ in (\ref{saddd}) gives us the final
form of the saddle point condition, 
\begin{equation}
\left( 1-\frac{(\beta ^{\prime })^2\Omega ^2C^2}{16Q}\left( \frac{\eta
^{\prime }(\beta ^{\prime },\Omega ,Q)}Q+\frac{\beta ^{\prime }}4\right)
^{-2}\right) =\lim_{N\rightarrow \infty }\frac 1{2NQ}\sum_{l=2}\sum_m\left( 
\frac{\eta ^{\prime }(\beta ^{\prime },\Omega ,Q)}Q+\frac{\beta ^{\prime }}{%
2l(l+1)}\right) ^{-1}  \label{sadsad}
\end{equation}
The alternate saddle point - if it exists - satisfies 
\begin{equation}
\eta ^{\prime }(\beta ^{\prime },\Omega ,Q)>\eta (\beta ^{\prime },Q)
\label{dd}
\end{equation}
because the LHS (\ref{sadsad}) must satisfy the pair of inequalities, 
\begin{equation}
0<\left( 1-\frac{(\beta ^{\prime })^2\Omega ^2C^2}{16Q}\left( \frac{\eta
^{\prime }(\beta ^{\prime },\Omega ,Q)}Q+\frac{\beta ^{\prime }}4\right)
^{-2}\right) <1.  \label{lhss}
\end{equation}
A useful condition that is equivalent to the upper bound is

\begin{equation}
\left( \frac{\beta ^{\prime }}4\right) ^2\frac{\Omega ^2C^2}Q\left( \frac{%
\eta ^{\prime }(\beta ^{\prime },\Omega ,Q)}Q+\frac{\beta ^{\prime }}%
4\right) ^{-1}<\left( \frac{\eta ^{\prime }(\beta ^{\prime },\Omega ,Q)}Q+%
\frac{\beta ^{\prime }}4\right) .  \label{dodi}
\end{equation}

From this we deduce that when the planetary spin $\Omega $ $>$ $\Omega
_c\equiv \sqrt{Q/C^2}$, the LHS(\ref{sadsad}) can be made to satisfy the
lower bound in (\ref{lhss}) by choosing $\eta ^{\prime }>\eta _c^{\prime
}(\beta ^{\prime },\Omega ,Q)>0$ where LHS(\ref{sadsad}) equals zero at $%
\eta _c^{\prime }(\beta ^{\prime },\Omega ,Q)<\infty $. LHS(\ref{sadsad})
equals one at $\eta ^{\prime }=\infty .$ If $\Omega $ $=$ $\Omega _c,$ then
clearly, $\eta _c^{\prime }(\beta ^{\prime },\Omega ,Q)=0$. But for $\Omega $
$<$ $\Omega _c,$ any $\eta ^{\prime }>0$ will satisfy the bounds in (\ref
{lhss}).

$\Omega $ $>$ $\Omega _c:$ RHS(\ref{sadsad}) equals one at the disordered
phase saddle point $\eta (\beta ^{\prime },Q)$ which is independent of $%
\Omega $ but equals $R(\eta _c^{\prime }(\beta ^{\prime },\Omega ,Q))$ at $%
\eta _c^{\prime }(\beta ^{\prime },\Omega ,Q)>0.$ The RHS(\ref{sadsad})
decreases to zero from the value $R(\eta _c^{\prime }(\beta ^{\prime
},\Omega ,Q))$ while LHS(\ref{sadsad}) increases to one from zero as $\eta
^{\prime }>\eta _c^{\prime }(\beta ^{\prime },\Omega ,Q)$ increases towards $%
\infty .$ There is an alternate (pure ground mode) saddle point solution of (%
\ref{sadsad}) since LHS(\ref{sadsad}) equals RHS(\ref{sadsad}) for some $%
\eta ^{\prime }\in (\eta _c^{\prime }(\beta ^{\prime },\Omega ,Q),$ $\infty
) $ by continuity.

On the other hand, if the planetary spin $\Omega $ is smaller than the
critical value $\Omega _c=\sqrt{Q/C^2}$, then at 
\[
\eta ^{\prime }(\beta ^{\prime },\Omega ,Q)=\eta (\beta ^{\prime },Q), 
\]
the RHS(\ref{sadsad}) equals one and the LHS(\ref{sadsad}) equals $L_c(\eta
(\beta ^{\prime },Q),\Omega )$ $\in (0,1)$. RHS(\ref{sadsad}) decreases from
one to zero as $\eta ^{\prime }$ increases from $\eta (\beta ^{\prime },Q)$
to $\infty .$ LHS(\ref{sadsad}) increases from $L_c(\eta (\beta ^{\prime
},Q),\Omega )$ towards one as $\eta ^{\prime }$ increases from $\eta (\beta
^{\prime },Q)$ to $\infty .$ There is again an alternate (pure ground mode)
saddle point solution of (\ref{sadsad}) since LHS(\ref{sadsad}) equals RHS(%
\ref{sadsad}) for some $\eta ^{\prime }\in (\eta (\beta ^{\prime },Q),Q),$ $%
\infty )$ by continuity.

This also shows that the thermodynamic limit exists for the spherical
Heisenberg models $H_H^N$ for all positive temperatures, in the sense that
for any $Q>0,$ $\Omega >0$ and all $\beta ^{\prime }>0,$ the saddle point
condition (\ref{saddd}) is well-defined and finite along the saddle points $%
\eta (\beta ^{\prime },Q)$ and $\eta ^{\prime }(\beta ^{\prime },\Omega ,Q)$.

It remains to show that the disordered phase is preferred at high positive
temperatures and the pure ground mode phase with counter-rotation $\alpha
_{10}<0$ is preferred at low positive temperatures. To compare the free
energy per site of these two phases we ignore for the moment the infinite
sum of logarithmic terms in $F$ and focus on the part which depends on $%
\alpha _{10}:$ 
\begin{eqnarray*}
\chi (\alpha _{10};\beta ,Q,\Omega ) &=&\eta ^{\prime }(\beta ^{\prime
})\left( 1-\frac{\alpha _{10}^2}Q\right) -\frac{\beta ^{\prime }}4\left(
\alpha _{10}^2+2\Omega C\alpha _{10}\right) \\
&=&\eta ^{\prime }(\beta ^{\prime })\left( 1-\frac{(\beta ^{\prime
})^2\Omega ^2C^2}{16Q}\left( \frac{\eta ^{\prime }(\beta ^{\prime },\Omega
,Q)}Q+\frac{\beta ^{\prime }}4\right) ^{-2}\right) \\
&&-\frac{\beta ^{\prime }}4\left[ \frac{(\beta ^{\prime })^2\Omega ^2C^2}{16}%
\left( \frac{\eta ^{\prime }(\beta ^{\prime },\Omega ,Q)}Q+\frac{\beta
^{\prime }}4\right) ^{-2}-\frac{\beta ^{\prime }\Omega ^2C^2}2\left( \frac{%
\eta ^{\prime }(\beta ^{\prime },\Omega ,Q)}Q+\frac{\beta ^{\prime }}%
4\right) ^{-1}\right]
\end{eqnarray*}
\begin{eqnarray*}
&=&-\frac{(\beta ^{\prime })^2\Omega ^2C^2}{16}\left( \frac{\eta ^{\prime
}(\beta ^{\prime },\Omega ,Q)}Q+\frac{\beta ^{\prime }}4\right) ^{-2}\left( 
\frac{\eta ^{\prime }}Q+\frac{\beta ^{\prime }}4\right) +\frac{(\beta
^{\prime })^2\Omega ^2C^2}8\left( \frac{\eta ^{\prime }(\beta ^{\prime
},\Omega ,Q)}Q+\frac{\beta ^{\prime }}4\right) ^{-1} \\
&=&\frac{(\beta ^{\prime })^2\Omega ^2C^2}{16}\left( \frac{\eta ^{\prime
}(\beta ^{\prime },\Omega ,Q)}Q+\frac{\beta ^{\prime }}4\right) ^{-1}.
\end{eqnarray*}
The same quantity for the disordered phase is given by 
\[
\chi (\alpha _{10}=0;\beta ,Q,\Omega )=\eta (\beta ^{\prime }). 
\]
Comparing them we get the following inequality which implies that the pure
ground mode phase is preferred: 
\begin{equation}
\frac{(\beta ^{\prime })^2\Omega ^2C^2}{16Q}>\frac{\eta (\beta ^{\prime },Q)}%
Q\left( \frac{\eta ^{\prime }(\beta ^{\prime },\Omega ,Q)}Q+\frac{\beta
^{\prime }}4\right) .  \label{compp}
\end{equation}

Fixing $\frac{\Omega ^2C^2}Q>0$, we deduce from (\ref{dd}) that (\ref{compp}%
) holds only if $\beta ^{\prime }$ $>\beta _{cc}^{\prime }(\Omega ,Q)$ where 
\[
\frac{(\beta _{cc}^{\prime })^2\Omega ^2C^2}{16Q}=\frac{\eta (\beta
_{cc}^{\prime },Q)}Q\left( \frac{\eta ^{\prime }(\beta _{cc}^{\prime
},\Omega ,Q)}Q+\frac{\beta _{cc}^{\prime }}4\right) 
\]
where such a positive value $\beta _{cc}^{\prime }<\infty $ exists by virtue
of the mean value theorem because for $\beta ^{\prime }$ near zero, 
\[
\frac{(\beta ^{\prime })^2\Omega ^2C^2}{16Q}<\frac{\eta (\beta ^{\prime },Q)}%
Q\left( \frac{\eta ^{\prime }(\beta ^{\prime },\Omega ,Q)}Q+\frac{\beta
^{\prime }}4\right) 
\]
since $\eta (\beta ^{\prime },Q)$ increases as $\beta ^{\prime }$ decreases,
and for $\beta ^{\prime }$ very large, 
\[
\frac{(\beta ^{\prime })^2\Omega ^2C^2}{16Q}>\frac{\eta (\beta ^{\prime },Q)}%
Q\left( \frac{\eta ^{\prime }(\beta ^{\prime },\Omega ,Q)}Q+\frac{\beta
^{\prime }}4\right) 
\]
since $\eta (\beta ^{\prime },Q)$ decreases down to zero as $\beta ^{\prime
} $ increases to $\infty $. We used the fact that both saddle points $\eta
(\beta ^{\prime },Q)$ and $\eta ^{\prime }(\beta ^{\prime },\Omega ,Q)$ are
smooth functions of $\beta ^{\prime }$ in the range $(0,\infty ).$

Returning to the infinite sum of logarithmic terms in $F,$ 
\[
-\lim_{N\rightarrow \infty }\frac 1{2N}\sum_{l=2}^{\sqrt{N}}\sum_{m=-l}^l\ln
\left( \frac{N\eta }Q+\frac{\beta ^{\prime }N}2\lambda _{lm}\right) , 
\]
we note that the value of this convergent sum for saddle point $\eta (\beta
^{\prime },Q)$ is bigger than that for $\eta ^{\prime }(\beta ^{\prime
},\Omega ,Q)$ in view of (\ref{dd}) but this difference is logarithmic in
the difference $\eta ^{\prime }(\beta ^{\prime },\Omega ,Q)-\eta (\beta
^{\prime },Q)>0,$ and is therefore dominated by the algebraic difference 
\[
\chi (\alpha _{10};\beta ,Q,\Omega )-\chi (\alpha _{10}=0;\beta ,Q) 
\]
discussed above.

This completes the proof that the disordered phase is preferred at high
positive temperatures but the ordered phase is preferred at low enough
temperatures where, unlike the negative critical point, the threshold value $%
\beta _{cc}^{\prime }(\Omega ,Q)$ depends on both relative enstrophy $Q$ and
planetary spin $\Omega .$ From (\ref{dodi}) we deduce that $\eta ^{\prime
}(\beta ^{\prime },\Omega ,Q)$ is linear in $\Omega $. Since $\eta (\beta
^{\prime },Q)$ does not depend on $\Omega ,$ this implies $\beta
_{cc}^{\prime }(\Omega ,Q)$ decreases as planetary spin $\Omega $ increases.
Thus, as $\Omega $ decreases to zero, the threshold value $\beta
_{cc}^{\prime }(\Omega ,Q)$ tends to $\infty ,$ and the disordered phase is
preferred at all positive temperatures in the case of a non-rotating massive
sphere.

Unlike the critical phenomenology of the barotropic fluid - sphere system at
very high energies (negative temperatures) which we have shown arises from
the reflection of the saddle point at the extreme value $\eta ^{*}$ (the
disordered phase does not satisfy the saddle point condition at negative $%
T^{\prime }$ when $|T^{\prime }|\ll 1),$ its critical phenomenology at
positive temperature is not so much based on the breakdown of the saddle
points as on the system's preference for a smaller free energy. Transitions
between these positive temperature phases for $\Omega >0$ are characterized
by a greater degree of smoothness than its negative temperature counterpart
since the free energy is automatically continuous at $\beta _{cc}^{\prime
}(\Omega ,Q)>0.$

\section{Proof of gapless and unique ground mode condensation}

An important result needed in the above exact solution of the spherical
Heisenberg model is the number and type of modes in the condensed phase. We
will prove that there is only one thermodynamically stable class of modes $%
(l=1)$ that has nonzero energy in the condensed phase and thence, exactly
these three modes are non-ergodic in this problem. The proof is based on the
existence of multiple saddle points $\eta ^{\prime }(\beta ^{\prime })\neq
\eta (\beta ^{\prime })$ and an application of Planck's theorem.

The first two parts of the proof are relatively short. The complete proof
will be given in the order: (1) at positive temperatures, the only
nontrivial condensed mode is associated with $\alpha _{10}<0$ signifying a
counter rotating solid body flow, (2) at negative temperatures, there cannot
be gapped nontrivial condensed modes where gapped means that there are
ergodic modes $l$ in between condensed modes $l^{\prime },$ and (3) at
negative temperatures, the only relevant nontrivial condensed mode is the
ground mode $\alpha _{10}>0$ which is associated with super-rotating
solid-body flows. Part (3) is longer because of the important property that
there is more than one saddle points at some negative $T$, namely (i) the
pure ground mode saddle point $\alpha _{10}>0$ and (ii) the condensed phase $%
\alpha _{2m}\neq 0$, $\alpha _{10}<0.$ We will show that saddle point (i)
has higher free energy per site than saddle point (ii) for all values of $%
T<T_c<0$ where the latter is condensed. By the extension of Planck's theorem
to negative temperatures, the preferred macrostate is the one with highest
free energy, namely the pure ground mode.

\subsection{(1) condensed modes at positive $T$ cannot have wavenumber $l>1$}

Assuming that there is a set of condensed modes with single $l>1$ at
positive $T,$ the finite part of the per site free energy expression $F$ -
after dropping the infinite sum - has the form 
\begin{eqnarray*}
\chi (\eta (\beta ^{\prime }),Q,\beta ^{\prime }) &=&\eta (\beta ^{\prime
})\left[ 1-\frac 1Q\left\{ \sum_{m=-l}^l\alpha _{lm}^2+\sum_{m=-1}^1\alpha
_{1m}^2\right\} \right] \\
-\frac{\beta ^{\prime }}4\sum_{m=-1}^1 &&\alpha _{1m}^2-\frac{\beta ^{\prime
}}2\Omega C\alpha _{10}-\left( \frac{\beta ^{\prime }\lambda _{lm}}2\right)
\sum_{m=-l}^l\alpha _{lm}^2.
\end{eqnarray*}
Similar to the approach in the previous section, the amplitudes $\alpha
_{lm} $ of these modes are fixed by additional equations of state 
\[
\frac{\partial F}{\partial \alpha _{lm}}=\frac{\partial \chi }{\partial
\alpha _{lm}}=-\left( \frac{\eta (\beta ^{\prime })}Q+\frac{\beta ^{\prime }%
}{l(l+1)}\right) \alpha _{lm}=0. 
\]
In order for at least one $\alpha _{lm}$ to be nonzero, 
\[
\left( \frac{\eta (\beta ^{\prime })}Q+\frac{\beta ^{\prime }}{l(l+1)}%
\right) =0 
\]
which contradicts the positivity of $\beta ^{\prime }$ and $\frac{\eta
(\beta ^{\prime })}Q.$ The distinguished ground mode $\alpha _{10}$ can be
nonzero because of the inhomogeneous term $-\frac{\beta ^{\prime }}2\Omega C$
in the corresponding equation of state for the amplitude.

\subsection{(2) condensed modes must be gapless at negative $T$}

We will prove next that at negative temperatures, there cannot be any
nonzero condensed modes $l^{\prime }\geq 2$ that is separated by ergodic
modes in between. The situation is clear from the simplest case of a single
additional nonzero condensed mode with $l^{\prime }>2$. Then, $l=2$
corresponds to ergodic modes which have to be integrated in the Gaussian
integrals, but similar to the above analysis of the equations of state, a
nonzero amplitude $\alpha _{l^{\prime }m}\neq 0$ implies that $\frac{\beta
^{\prime }\lambda _{l^{\prime }m}}2+\frac{\eta (\beta ^{\prime })}Q=0$ which
in turn means that $\frac{\beta ^{\prime }\lambda _{2m}}2+\frac{\eta (\beta
^{\prime })}Q<0,$ violating the Gaussian integrability condition.

\subsection{(3) the only condensed mode at negative $T$ is $\alpha _{10}>0$}

Assume (for reductio ad absurdum) that more than one class of modes have
nonzero amplitude in the condensed phase of this problem, namely those
belonging to wavenumbers $l_1,l_2$ and also $l=1$ (this being the
distinguished ground mode of the problem, has to be part of the condensed
phase) where $l_2=l_1+1=3$. Other cases where $l_2,$ $l_1$ and $l=1$ are not
consecutive, cannot appear in the condensed phase of a solvable spherical
model, because the Gaussian solvability conditions 
\[
\frac{\beta ^{\prime }\lambda _{lm}}2+\frac \eta Q=\frac{\beta ^{\prime }}{%
2l(l+1)}+\frac \eta Q>0 
\]
are violated as shown above.

Then, the restricted partition function is given by 
\[
Z_H^N(\beta ,Q,\Omega )\propto \int \prod_{m=-1}^1d\alpha
_{1m}\prod_{m=-l_1}^{l_1}d\alpha _{l_1m}\text{ }\prod_{m=-l_2}^{l_2}d\alpha
_{l_2m}Z_H^N(\alpha _{1m},\alpha _{l_1m},\alpha _{l_2m};\beta ,Q,\Omega ) 
\]
\[
=\int \prod_{m=-1}^1d\alpha _{1m}\prod_{m=-l_1}^{l_1}d\alpha _{l_1m}\text{ }%
\prod_{m=-l_2}^{l_2}d\alpha _{l_2m}\int_{a-i\infty }^{a+i\infty }\frac{d\eta 
}{2\pi i}\exp \left\{ N\left[ 
\begin{array}{c}
\eta -\left( \frac{\beta \lambda _{l_1m}}{2N}+\frac \eta Q\right)
\sum_{m=-1}^1\alpha _{l_1m}^2 \\ 
-\left( \frac{\beta \lambda _{l_2m}}{2N}+\frac \eta Q\right)
\sum_{m=-1}^1\alpha _{l_2m}^2 \\ 
-\left( \frac \beta {4N}+\frac \eta Q\right) \sum_{m=-1}^1\alpha _{1m}^2 \\ 
-\frac \beta {2N}\Omega C\alpha _{10}
\end{array}
\right] \right\} 
\]
\[
\int D_{l\neq 1,l_1,l_2}(\alpha )\exp \left( -N\sum_{l\neq 1,l_1,l_2}^\infty
\sum_{m=-l}^l\left( \frac \beta {2N}\lambda _{lm}+\frac \eta Q\right) \alpha
_{lm}^2\right) . 
\]
Standard Gaussian integration is used to evaluate the last integral, which
yields, after scaling $\beta ^{\prime }N=\beta ,$ 
\[
\int_{l\neq 1,l_1,l_2}D(\alpha )\exp \left( -\sum_{l\neq
1,l_1,l_2}\sum_{m=-l}^l\left( \frac{\beta ^{\prime }N\lambda _{lm}}2+\frac{%
N\eta }Q\right) \alpha _{lm}^2\right) =\prod_{l\neq 1,l_1,l_2}^{\sqrt{N}%
}\prod_{m=-l}^l\left( \frac \pi {\frac{N\eta }Q+\frac{\beta ^{\prime }N}%
2\lambda _{lm}}\right) ^{1/2}, 
\]
provided the physically significant Gaussian conditions hold: for $l\neq
1,l_1,l_2,$ 
\[
\frac{\beta ^{\prime }\lambda _{lm}}2+\frac \eta Q=\frac{\beta ^{\prime }}{%
2l(l+1)}+\frac \eta Q>0. 
\]
Then the restricted partition function takes the form 
\[
Z_H^N(\alpha _{1m},\alpha _{l_1m},\alpha _{l_2m};\beta ,Q,\Omega )\propto
\frac 1{2\pi i}\int_{a-i\infty }^{a+i\infty }d\eta \text{ }\exp \left\{
N\left[ 
\begin{array}{c}
\eta -\left\{ 
\begin{array}{c}
\left( \frac{\beta ^{\prime }\lambda _{l_1m}}2+\frac \eta Q\right)
\sum_{m=-1}^1\alpha _{l_1m}^2 \\ 
+\left( \frac{\beta ^{\prime }\lambda _{l_2m}}2+\frac \eta Q\right)
\sum_{m=-1}^1\alpha _{l_2m}^2
\end{array}
\right\} \\ 
-\left( \frac{\beta ^{\prime }}4+\frac \eta Q\right) \sum_{m=-1}^1\alpha
_{1m}^2-\frac{\beta ^{\prime }}2\Omega C\alpha _{10} \\ 
-\frac 1{2N}\sum_{l\neq 1,l_1,l_2}\sum_{m=-l}^l\ln \left( \frac{N\eta }Q+%
\frac{\beta ^{\prime }N}2\lambda _{lm}\right)
\end{array}
\right] \right\} 
\]
where the free energy per site evaluated at the most probable macrostate is $%
-\frac 1{\beta ^{\prime }}$ $F(\eta (\beta ^{\prime }),Q,\beta ^{\prime })\,$
with 
\begin{eqnarray*}
F(\eta (\beta ^{\prime }),Q,\beta ^{\prime }) &=&\eta (\beta ^{\prime
})\left[ 1-\frac 1Q\left\{ \sum_{m=-l_1}^{l_1}\alpha
_{l_1m}^2+\sum_{m=-l_2}^{l_2}\alpha _{l_2m}^2+\sum_{m=-1}^1\alpha
_{1m}^2\right\} \right] \\
-\frac{\beta ^{\prime }}4\sum_{m=-1}^1 &&\alpha _{1m}^2-\frac{\beta ^{\prime
}}2\Omega C\alpha _{10}-\left\{ \left( \frac{\beta ^{\prime }\lambda _{l_1m}}%
2\right) \sum_{m=-l_1}^{l_1}\alpha _{l_1m}^2+\left( \frac{\beta ^{\prime
}\lambda _{l_2m}}2\right) \sum_{m=-l_1}^{l_2}\alpha _{l_2m}^2\right\}
\end{eqnarray*}
\[
-\frac 1{2N}\sum_{l\neq 1,l_1,l_2}\sum_m\ln \left( \frac{N\eta }Q+\frac{%
\beta ^{\prime }N}2\lambda _{lm}\right) . 
\]

The saddle point condition gives one equation for the determination of the
variables $\eta ,\alpha _{1m},$ $\alpha _{l_1m},$ $\alpha _{l_1m}$ in terms
of inverse temperature $\beta ^{\prime }$, relative enstrophy $Q,$ and the
fixed rate of spin $\Omega >0$ of the planetary frame, 
\[
0=\frac{\partial F}{\partial \eta }=1-\frac 1Q\left\{
\sum_{m=-l_1}^{l_1}\alpha _{l_1m}^2+\sum_{m=-l_2}^{l_2}\alpha
_{l_2m}^2+\sum_{m=-1}^1\alpha _{1m}^2\right\} -\frac 1{2NQ}\sum_{l\neq
1,l_1,l_2}\sum_m\left( \frac{\eta (\beta ^{\prime })}Q+\frac{\beta ^{\prime }%
}2\lambda _{lm}\right) ^{-1} 
\]
where $\eta =\eta (\beta ^{\prime })$ is taken to be the value of the saddle
point. Equations to close the system are provided by equations of state (or
Planck's theorem) for the condensed phase: for $l=1,$ 
\begin{eqnarray*}
0 &=&\frac{\partial F}{\partial \alpha _{10}}=-\left( \frac{2\eta (\beta
^{\prime })}Q+\frac{\beta ^{\prime }}2\right) \alpha _{10}-\frac{\beta
^{\prime }}2\Omega C \\
0 &=&\frac{\partial F}{\partial \alpha _{1,\pm 1}}=-\left( \frac{2\eta
(\beta ^{\prime })}Q+\frac{\beta ^{\prime }}2\right) \alpha _{1,\pm 1};
\end{eqnarray*}
and for $l=l_1,l_2,$ $m=-l,...,0,...,l,$%
\begin{eqnarray*}
0 &=&\frac{\partial F}{\partial \alpha _{l_1m}}=-\left( \frac{2\eta (\beta
^{\prime })}Q+\beta ^{\prime }\lambda _{l_1m}\right) \alpha _{l_1m} \\
0 &=&\frac{\partial F}{\partial \alpha _{l_2m}}=-\left( \frac{2\eta (\beta
^{\prime })}Q+\beta ^{\prime }\lambda _{l_2m}\right) \alpha _{l_2m}.
\end{eqnarray*}
We assume that $\alpha _{1m}\neq 0$ for some $m=-1,0,1.$

The proof by contradiction continues by further assuming that at least one
of the $l^{\prime }=l_1,l_2$ equations of state are satisfied by nonzero
amplitudes, say $\alpha _{l_1m}\neq 0$ for some $m=-l_1,...,0,...,l_1.$ Then 
\[
\left( \frac{2\eta (\beta ^{\prime })}Q+\beta ^{\prime }\lambda
_{l_1m}\right) =0 
\]
which in turn implies 
\begin{eqnarray*}
\left( \frac{2\eta (\beta ^{\prime })}Q+\beta ^{\prime }\lambda
_{l_2m}\right) &>&0, \\
\left( \frac{2\eta (\beta ^{\prime })}Q+\frac{\beta ^{\prime }}2\right) &<&0,
\end{eqnarray*}
and thence, 
\begin{eqnarray*}
\alpha _{l_2m} &=&0, \\
\alpha _{1,\pm 1} &=&0, \\
\alpha _{10} &\neq &0.
\end{eqnarray*}
The first of the $l=1$ equations of state then implies 
\begin{equation}
\alpha _{10}=-\frac{\beta ^{\prime }}2\Omega C\left( \frac{2\eta (\beta
^{\prime })}Q+\frac{\beta ^{\prime }}2\right) ^{-1}<0\text{ for }\beta
^{\prime }<0  \label{altt}
\end{equation}
since $\Omega C>0;$ and vice-versa for $\beta ^{\prime }>0.$

The contradiction at negative temperatures is obtained first by showing that
although the saddle point equation 
\begin{equation}
1-\frac 1Q\left\{ \sum_{m=-2}^2\alpha _{2m}^2+\alpha _{10}^2\right\}
=\lim_{N\rightarrow \infty }\frac 1{2NQ}\sum_{l\neq 1,2,3}\sum_m\left( \frac{%
\eta (\beta ^{\prime })}Q+\frac{\beta ^{\prime }}2\lambda _{lm}\right) ^{-1}
\label{sud}
\end{equation}
has solution $\eta ^{\prime }(\beta ^{\prime }),$ $\alpha _{10}<0,$ $\alpha
_{2m}\neq 0$ in addition to the pure ground mode solution $\eta (\beta
^{\prime })$, $\alpha _{10}>0,$ $\alpha _{2m}=0$ at large values of the
scaled inverse temperature $\beta ^{\prime }<\beta _c^{\prime }<0$ $,$ these
counter-rotating solutions have lower values of $F^{\prime }=F(\eta ^{\prime
}(\beta ^{\prime }),Q,\beta ^{\prime })$ than the same expression $F$ for
the pure ground mode solution$.$ Here $\beta _c^{\prime }<0$ is the most
negative inverse temperature for which (\ref{sud}) has unique saddle point,
namely the disordered phase, $\eta ,\alpha _{10}=0,\alpha _{2m}=0.$ This
critical point has the same value critical inverse temperature as obtained
in the last section because the RHS (\ref{sud}) is the same as its
counterpart in the pure ground mode case in the thermodynamic limit. For all 
$\beta ^{\prime }<\beta _c^{\prime },$ an argument similar to that used in
the previous section to prove the existence of the pure ground mode saddle
point, can be used here to prove the alternative saddle point equation (\ref
{sud}) has two saddle points, namely the sticking one, 
\begin{equation}
\eta ^{\prime }(\beta ^{\prime })=-\frac{\beta ^{\prime }Q}{12},\alpha
_{10}<0,\alpha _{2m}\neq 0  \label{comp}
\end{equation}
and reflected pure ground mode solution 
\begin{equation}
\eta (\beta ^{\prime })>\eta ^{*}=-\frac{\beta ^{\prime }Q}4,\alpha
_{10}>0,\alpha _{2m}=0.  \label{ref1}
\end{equation}

By the extension of Planck's theorem to negative temperatures, solutions $%
\alpha _{10}<0,$ $\alpha _{2m}\neq 0$ at $\beta ^{\prime }<0$ are not the
most probable and statistically stable macrostate because they have lower
per site free energy.

Using the sticking property, we deduce that the free energy expression $%
F(\eta ^{\prime }(\beta ^{\prime }),Q,\beta ^{\prime })$ is given by 
\begin{eqnarray*}
F^{\prime } &=&\eta ^{\prime }(\beta ^{\prime })-\frac{\beta ^{\prime }}%
2\Omega C\alpha _{10}-\left( \frac{\eta ^{\prime }(\beta ^{\prime })}Q+\frac{%
\beta ^{\prime }}4\right) \alpha _{10}^2 \\
-\lim_{N\rightarrow \infty } &&\frac 1{2N}\sum_{l\neq 1,2,3}\sum \ln \left( 
\frac{N\eta ^{\prime }}Q+\frac{\beta ^{\prime }N}2\lambda _{lm}\right) ,
\end{eqnarray*}
which apart from the first terms in the infinite sum that vanish like $%
N^{-1}\ln N$ for large $N,$ and the new value $\eta ^{\prime }(\beta
^{\prime })<\eta (\beta ^{\prime })$, has the same form as $F(\eta (\beta
^{\prime }),Q,\beta ^{\prime })$ in the pure ground mode case. Equation (\ref
{comp}) and $\alpha _{10}<0$ imply that the first two terms of $F^{\prime }$
is smaller than those in $F$ and the third term is bigger. So for $\Omega >0$
large enough, since $\eta ^{\prime }(\beta ^{\prime })$ is independent of $%
\Omega ,$ we can make $F^{\prime }<F.$

We can do more by using the sticking property of $\eta ^{\prime }(\beta
^{\prime }).$ Substituting $\eta ^{\prime }(\beta ^{\prime })=-\frac{\beta
^{\prime }Q}{12}$ and $\alpha _{10}<0$ from (\ref{altt}) into the first
three terms in $F^{\prime },$ and using the reflection property of the
saddle point $\eta (\beta ^{\prime }),$ namely, $\left( \frac{2\eta (\beta
^{\prime })}Q+\frac{\beta ^{\prime }}2\right) >0,$ we obtain the required
comparison at $\beta ^{\prime }<0,$ that is, 
\begin{eqnarray*}
F_3^{\prime }(\eta ^{\prime }(\beta ^{\prime })) &=&-\frac{\beta ^{\prime }Q%
}{12}+\frac 12\left( \frac{\beta ^{\prime }}2\right) ^2\Omega ^2C^2\left( 
\frac{\beta ^{\prime }}3\right) ^{-1} \\
&=&-\frac{\beta ^{\prime }Q}{12}+\frac{3\beta ^{\prime }}8\Omega ^2C^2<-%
\frac{\beta ^{\prime }Q}4 \\
&<&-\frac{\beta ^{\prime }Q}4+\frac 12\left( \frac{\beta ^{\prime }}2\right)
^2\Omega ^2C^2\left( \frac{2\eta (\beta ^{\prime })}Q+\frac{\beta ^{\prime }}%
2\right) ^{-1} \\
&<&F_3(\eta (\beta ^{\prime })).
\end{eqnarray*}
In other words, for all $\beta ^{\prime }<\beta _c^{\prime }$ the per site
free energy expression $F(\eta (\beta ^{\prime }))>F^{\prime }(\eta ^{\prime
}(\beta ^{\prime })).$ Thus, the pure ground mode condensed phase is
preferred in the thermodynamic limit.

This completes the proof of the result that for all values of planetary spin
at negative temperatures, there is exactly one thermodynamically stable
saddle point, that is the disordered phase, $\eta ,\alpha _{10}=0$ at
negative $\beta ^{\prime }>\beta _c^{\prime }$, and the pure ground mode
condensed phase, $\eta ,\alpha _{10}>0$ at negative $\beta ^{\prime }<\beta
_c^{\prime }.$

An important consequence of this result is that the alternative condensed
phase, $\eta ^{\prime }(\beta ^{\prime }),$ $\alpha _{10}<0,$ $\alpha
_{2m}\neq 0$ can in principle be thermodynamically stable in other
geophysical flow problems such as the generalized Shallow Water Equations
(GSWE) on a massive rotating sphere.

\section{Solution of the spherical Ising model for $\Omega =0$}

We briefly review the exact solution of the spherical Ising model in the
special case of a fixed frame and refer the reader to the literature for
details \cite{Lim06a}. In view of the sections on the role of energy and
angular momentum in the formulation of a statistical mechanics for the
barotropic fluid - solid sphere system, one might ask why we need to solve
this case when we have already solved the general $\Omega >0$ case in the
previous section. This case applies to the situation of a non-rotating
infinitely massive solid sphere which behaves like three infinite reservoirs
of angular momentum for each of the three principle directions. As will be
shown below this case differs from the $\Omega >0$ case because without a
distinguished axis of rotation, the ground or ordered modes in the problem
has $SO(3)$ degeneracy (due to the 3-fold degeneracy of the $l=1$ spherical
harmonics). Symmetry breaking in the phase transition for this case is
therefore more explicit than in the $\Omega >0$ case. In fact it is clearly
an instance of so-called spontaneous symmetry breaking (SSB) in the general
Ginsburg-Landau formulation of second order phase transitions. Similar to
the Heisenberg models, the negative critical temperature here is due to the
anti-ferromagnetic nature of the logarithmic interaction in the energy.

The partition function for the spherical Ising model has the form 
\[
Z_N\propto \int D(\vec{s})\exp \left( -\beta H_N(\vec{s})\right) \delta
\left( Q\frac N{4\pi }-\sum_{j=1}^N\vec{s}_j\cdot \vec{s}_j\right) 
\]
where the path-integral is taken over all microstates $\vec{s}$ with zero
circulation. In the thermodynamic or continuum limit as $N\rightarrow \infty
,$ the partition function is calculated using Laplace's integral form, 
\begin{eqnarray*}
Z_N &\propto &\int D(\vec{s})\exp \left( -\beta H_N(\vec{s})\right) \delta
\left( Q\frac N{4\pi }-\sum_{j=1}^N\vec{s}_j\cdot \vec{s}_j\right) \\
&=&\int D(\vec{s})\exp \left( -\beta H_N(\vec{s})\right) \left( \frac 1{2\pi
i}\int_{a-i\infty }^{a+i\infty }d\eta \exp \left( \eta \left( Q\frac N{4\pi
}-\sum_{j=1}^N\vec{s}_j\cdot \vec{s}_j\right) \right) \right) .
\end{eqnarray*}
Solution of the Gaussian integrals require diagonalizing the interaction in $%
H_N$ in terms of the spherical harmonics $\left\{ \psi _{lm}\right\}
_{l=1,}^\infty $ which are natural Fourier modes for Laplacian eigenvalue
problems on $S^2$ with zero circulation. Since the ordered modes are
associated with $l=1,$ we do not need to integrate over $\alpha _{1m}$ and
henceforth discuss only the restricted partition function 
\[
Z_N(\beta ,Q;\alpha _{10},\alpha _{1,\pm 1})\propto \int D_{l\geq 2}(\alpha
)\exp \left( -\frac \beta 2\sum_{l=1}\sum_{m=-l}^l\lambda _{lm}\alpha
_{lm}^2\right) 
\]
\[
\left( \frac 1{2\pi i}\int_{a-i\infty }^{a+i\infty }d\eta \exp \left( \eta
N\left( 1-\frac{4\pi }Q\sum_{l=1}\sum_{m=-l}^l\alpha _{lm}^2\right) \right)
\right) 
\]
where the eigenvalues of the Green's function for the Laplace-Beltrami
operator on $S^2$ are 
\[
\lambda _{lm}=\frac 1{l(l+1)},\text{ }l=1,...,\sqrt{N},\text{ }%
m=-l,...,0,...,l 
\]
and $\alpha _{lm}$ are the corresponding amplitudes.

Next we exchange the order of integration, which is allowed provided $a>0$
is chosen large enough so that the integrand is absolutely convergent, and
rescale $\beta ^{\prime }N=\beta ,$%
\[
Z_N(\beta ,Q;\alpha _{10},\alpha _{1,\pm 1})\propto \frac 1{2\pi
i}\int_{a-i\infty }^{a+i\infty }d\eta \exp \left( \eta N\left( 1-\frac{4\pi }%
Q\sum_{m=-l}^l\alpha _{1m}^2\right) -\frac{\beta ^{\prime }N}%
2\sum_{m=-1}^1\lambda _{1m}\alpha _{1m}^2\right) 
\]
\[
\int_{l\geq 2}D(\alpha )\exp \left( -\sum_{l=2}\sum_{m=-l}^l\left( \frac{%
\beta ^{\prime }N\lambda _{lm}}2+N\eta \frac{4\pi }Q\right) \alpha
_{lm}^2\right) . 
\]
We stress that 
\[
\exp \left( -\frac{\beta ^{\prime }N}2\sum_{m=-1}^1\lambda _{1m}\alpha
_{1m}^2\right) 
\]
is not part of the integrand since the ground modes are not integrated.
Moreover, by standard procedures for a Gaussian, we explicitly solve the
inner integral$,$ 
\[
\int_{l\geq 2}D(\alpha )\exp \left( -\sum_{l=2}\sum_{m=-l}^l\left( \frac{%
\beta ^{\prime }N\lambda _{lm}}2+N\eta \frac{4\pi }Q\right) \alpha
_{lm}^2\right) =\prod_{l=2}^{\sqrt{N}}\prod_{m=-l}^l\left( \frac \pi {N\eta 
\frac{4\pi }Q+\frac{\beta ^{\prime }N}2\lambda _{lm}}\right) ^{1/2}, 
\]
provided the following physically important conditions hold 
\begin{equation}
\frac{\beta ^{\prime }\lambda _{lm}}2+\eta \frac{4\pi }Q>0,\text{ }l=2,...,%
\sqrt{N},\text{ }m=-l,...,0,...,l.  \label{pos1}
\end{equation}

Thus, 
\[
Z_N(\beta ,Q;\alpha _{10},\alpha _{1,\pm 1})\propto \int_{a-i\infty
}^{a+i\infty }d\eta \exp N\left[ 
\begin{array}{c}
\eta \left( 1-\frac{4\pi }Q\sum_{m=-1}^1\alpha _{1m}^2\right) -\frac{\beta
^{\prime }}2\sum_{m=-1}^1\lambda _{1m}\alpha _{1m}^2 \\ 
-\frac 1{2N}\sum_{l=2}\sum_m\ln \left( N\eta \frac{4\pi }Q+\frac{\beta
^{\prime }N}2\lambda _{lm}\right)
\end{array}
\right] , 
\]
which can be written in steepest descent form, 
\begin{eqnarray*}
Z &\propto &\lim_{N\rightarrow \infty }\frac 1{2\pi i}\int_{a-i\infty
}^{a+i\infty }d\eta \exp \left( NF(\eta ,Q,\beta ^{\prime })\right) \\
&=&\frac 1{2\pi i}\int_{a-i\infty }^{a+i\infty }d\eta \exp \left( -\beta
^{\prime }g(\eta ,Q,\beta ^{\prime })\right) \\
&=&\exp \left( -\beta ^{\prime }g(\eta (\beta ^{\prime }),Q,\beta ^{\prime
})\right)
\end{eqnarray*}
in the thermodynamic limit as $N\rightarrow \infty ,$where the free energy
per site, after separating out the 3-fold degenerate ground states $\psi
_{10,}\psi _{l,\pm 1},$ is given by 
\begin{eqnarray*}
&&-\frac 1{\beta ^{\prime }}\text{ }F(\eta (\beta ^{\prime }),Q,\beta
^{\prime })\,\text{ with} \\
F(\eta (\beta ^{\prime }),Q,\beta ^{\prime }) &=&\eta (\beta ^{\prime
})\left( 1-\frac{4\pi }Q\sum_{m=-1}^1\alpha _{1m}^2\right) -\frac{\beta
^{\prime }}2\sum_{m=-1}^1\lambda _{1m}\alpha _{1m}^2 \\
-\frac 1{2N}\sum_{l=2} &&\sum_m\ln \left( N\eta (\beta ^{\prime })\frac{4\pi 
}Q+\frac{\beta ^{\prime }N}2\lambda _{lm}\right)
\end{eqnarray*}
and the total free energy 
\[
g(\eta (\beta ^{\prime }),Q,\beta ^{\prime })=\lim_{N\rightarrow \infty
}\left( -\frac N{\beta ^{\prime }}\text{ }F(\eta (\beta ^{\prime }),Q,\beta
^{\prime })\right) . 
\]
The saddle point parameter $\eta =\eta (\beta ^{\prime })$ is determined by
solving the following set of four equations for $(\eta ,$ $\alpha _{1m})$ in
terms of given values of inverse temperature $\beta ^{\prime }$ and relative
enstrophy $Q.$

The saddle point condition is 
\begin{equation}
0=\frac{\partial F}{\partial \eta }=\left( 1-\frac{4\pi }Q\sum_{m=-1}^1%
\alpha _{1m}^2\right) -\frac{2\pi }Q\sum_{l=2}^{\sqrt{N}}\sum_{m=-l}^l\left(
N\eta \frac{4\pi }Q+\frac{\beta ^{\prime }N}2\lambda _{lm}\right) ^{-1}.
\label{sad1}
\end{equation}
A set of three additional conditions to close the system is given by the
equations of state for $m=-1,0,1,$%
\begin{equation}
0=\frac{\partial F}{\partial \alpha _{1m}}=\left( \frac{8\pi \eta }Q+\beta
^{\prime }\lambda _{1m}\right) \alpha _{1m}.  \label{st1}
\end{equation}

The last 3 equations have solutions 
\[
\alpha _{1m}\text{ }=0\text{ or }\frac{8\pi \eta }Q+\beta ^{\prime }\lambda
_{1m}=0,\text{ for each }m. 
\]
This means that in order to have nonzero amplitudes in at least one of the
ground / condensed states (which are the only ones to have angular
momentum), $\frac{4\pi \eta ^{*}}Q=-\frac{\beta ^{\prime }}4,$which implies
that the inverse temperature must be negative, $\beta ^{\prime }<0.$

The Gaussian condition (\ref{pos1}) on the modes with $l=2,$%
\[
\frac{\beta ^{\prime }}{12}-\frac{\beta ^{\prime }}2>0, 
\]
can only be satisfied by $\beta ^{\prime }<0$ when there is any energy in
the angular momentum containing ground modes.

Substituting this nonzero solution into the saddle point equation yields 
\begin{eqnarray*}
0 &=&\left( 1-\frac{4\pi }Q\sum_{m=-1}^1\alpha _{1m}^2\right) -\frac{4\pi }Q%
\frac{T^{\prime }}N\sum_{l=2}^{\sqrt{N}}\sum_{m=-l}^l\left( \lambda
_{lm}-\frac 12\right) ^{-1} \\
&=&\left( 1-\frac{4\pi }Q\sum_{m=-1}^1\alpha _{1m}^2\right) -\frac{T^{\prime
}}{T_c^{^{\prime }}}
\end{eqnarray*}
where the critical inverse temperature is negative, finite, and inversely
proportional to the relative enstrophy $Q,$%
\[
-\infty <\beta _c^{\prime }=\frac{4\pi }{QN}\sum_{l=2}^{\sqrt{N}%
}\sum_{m=-l}^l\left( \lambda _{lm}-\frac 12\right) ^{-1}<0. 
\]
The saddle point equation provides a way to compute the equilibrium
amplitudes of the ground modes for temperatures hotter than $T_c^{^{\prime
}}<0,$ that is, 
\[
\text{ for }T_c^{^{\prime }}<T^{\prime }<0,\text{ }\sum_{m=-1}^1\alpha
_{1m}^2(T^{\prime })=\frac Q{4\pi }\left( 1-\frac{T^{\prime }}{T_c^{^{\prime
}}}\right) . 
\]

The above argument shows that at positive temperatures (low barotropic
energy), there cannot be any energy in the solid-body rotating modes. In
other words, there is no phase transition at positive temperatures when $%
\Omega =0$. This is the spin-lattice representation of the self-organization
of barotropic energy into a large-scale coherent flow at very high energies
in the form of symmetry-breaking Goldstone modes. The reader should compare
the predictions of the spherical model for barotropic vortex statistics
contained in these formulae with the results of Monte-Carlo simulations in
the next chapter. In particular, the linear dependence of the negative
critical temperature $T_c=T_c(Q)$ $<0$ on the relative enstrophy should be
noted in the spherical model solution as well as the Monte-Carlo simulations 
\cite{dinglim1}.

The free energy per site in the thermodynamic limit $(N\rightarrow \infty )$
has the form 
\[
f(\eta ,Q,\beta ^{\prime })=\lim_{N\rightarrow \infty }\left( -\frac 1{\beta
^{\prime }}\text{ }F(\eta (\beta ^{\prime }),Q,\beta ^{\prime })\right)
=u-Ts 
\]

\[
=\frac 14\sum_{m=-1}^1\alpha _{1m}^2+\frac Q{16\pi }\left( 1-\frac{4\pi }%
Q\sum_{m=-1}^1\alpha _{1m}^2\right) +\lim_{N\rightarrow \infty }\frac{%
T^{\prime }}{2N}\sum_{l=2}\sum_m\ln \frac N{2T^{\prime }}\left( -\frac
12+\lambda _{lm}\right) 
\]
\[
=\frac Q{16\pi }+\lim_{N\rightarrow \infty }\frac{T^{\prime }}{2N}%
\sum_{l=2}\sum_m\ln \frac N{2T^{\prime }}\left( -\frac 12+\lambda
_{lm}\right) 
\]
and 
\begin{eqnarray*}
\frac d{dT^{\prime }}f(\eta ^{*},Q,T) &=&\lim_{N\rightarrow \infty }\frac
1{2N}\sum_{l=2}\sum_m\ln \frac N{2T^{\prime }}\left( -\frac 12+\lambda
_{lm}\right) \\
-\lim_{N\rightarrow \infty } &&\frac{T^{\prime }}{2N}\sum_{l=2}\sum_m\frac{%
2T^{\prime }}N\left( -\frac 12+\lambda _{lm}\right) ^{-1}\frac N{2T^{\prime
2}}\left( -\frac 12+\lambda _{lm}\right) \\
&=&\lim_{N\rightarrow \infty }\frac 1{2N}\sum_{l=2}\sum_m\ln \frac
N{2T^{\prime }}\left( -\frac 12+\lambda _{lm}\right) -\lim_{N\rightarrow
\infty }\frac 1{2N}\sum_{l=2}\sum_m1 \\
&=&-1+\lim_{N\rightarrow \infty }\frac 1{2N}\sum_{l=2}\sum_m\ln \frac
N{2T^{\prime }}\left( -\frac 12+\lambda _{lm}\right)
\end{eqnarray*}
when $\sum_{m=-1}^1\alpha _{1m}^2>0$ and $T_c^{^{\prime }}<T^{\prime }<0.$
We used 
\[
\lim_{N\rightarrow \infty }\frac 1{2N}\sum_{l=2}\sum_m1=1 
\]
as well as the sticking value of the saddle point 
\[
\frac{4\pi \eta ^{*}}Q=-\frac{\beta ^{\prime }}4. 
\]

When $T^{\prime }<T_c^{^{\prime }}<0,$ the saddle point equation implies
that 
\begin{equation}
\lim_{N\rightarrow \infty }\frac{2\pi }{QN}\sum_{l=2}^{\sqrt{N}%
}\sum_{m=-l}^l\left( \eta \frac{4\pi }Q+\frac 1{2T^{\prime }}\lambda
_{lm}\right) ^{-1}=1  \label{sadd1}
\end{equation}
giving 
\begin{eqnarray*}
f(\eta (\beta ^{\prime }),Q,\beta ^{\prime }) &=&\lim_{N\rightarrow \infty
}\left( -\frac 1{\beta ^{\prime }}\text{ }F(\eta (\beta ^{\prime }),Q,\beta
^{\prime })\right) \\
&=&-\eta (T^{\prime })T^{\prime }+\lim \frac{T^{\prime }}{2N}%
\sum_{l=2}\sum_m\ln \left( N\eta (T^{\prime })\frac{4\pi }Q+\frac
N{2T^{\prime }}\lambda _{lm}\right)
\end{eqnarray*}
because $\sum_{m=-1}^1\alpha _{1m}^2=0$ and 
\begin{eqnarray*}
\frac d{dT^{\prime }}f(\eta (T^{\prime }),Q,T) &=&-\eta (T^{\prime })-\eta
^{\prime }(T^{\prime })T^{\prime }+\lim \frac 1{2N}\sum_{l=2}\sum_m\ln
\left( N\eta (T^{\prime })\frac{4\pi }Q+\frac N{2T^{\prime }}\lambda
_{lm}\right) \\
+\lim \frac{T^{\prime }}{2N}\sum_{l=2} &&\sum_m\left( \eta (T^{\prime })%
\frac{4\pi }Q+\frac 1{2T^{\prime }}\lambda _{lm}\right) ^{-1}\left( -\frac{%
\lambda _{lm}}{2(T^{\prime })^2}+\eta ^{\prime }(T^{\prime })\frac{4\pi }%
Q\right)
\end{eqnarray*}
\begin{eqnarray*}
&=&-\eta (T^{\prime })-\eta ^{\prime }(T^{\prime })T^{\prime }+\lim \frac
1{2N}\sum_{l=2}\sum_m\ln \left( N\eta (T^{\prime })\frac{4\pi }Q+\frac
N{2T^{\prime }}\lambda _{lm}\right) \\
+\lim \eta ^{\prime }(T^{\prime })T^{\prime }\frac{2\pi }{QN}\sum_{l=2}
&&\sum_m\left( \eta (T^{\prime })\frac{4\pi }Q+\frac 1{2T^{\prime }}\lambda
_{lm}\right) ^{-1} \\
-\lim \frac 1{4NT^{\prime }}\sum_{l=2} &&\sum_m\lambda _{lm}\left( \eta
(T^{\prime })\frac{4\pi }Q+\frac 1{2T^{\prime }}\lambda _{lm}\right) ^{-1}
\end{eqnarray*}
\begin{eqnarray*}
&=&-\eta (T^{\prime })-\eta ^{\prime }(T^{\prime })T^{\prime }+\lim \frac
1{2N}\sum_{l=2}\sum_m\ln \left( N\eta (T^{\prime })\frac{4\pi }Q+\frac
N{2T^{\prime }}\lambda _{lm}\right) \\
+\eta ^{\prime }(T^{\prime })T^{\prime }-\lim \frac 1{4NT^{\prime
}}\sum_{l=2} &&\sum_m\lambda _{lm}\left( \eta (T^{\prime })\frac{4\pi }%
Q+\frac 1{2T^{\prime }}\lambda _{lm}\right) ^{-1}
\end{eqnarray*}
\begin{eqnarray*}
&=&-\eta (T^{\prime })+\lim \frac 1{2N}\sum_{l=2}\sum_m\ln \left( N\eta
(T^{\prime })\frac{4\pi }Q+\frac N{2T^{\prime }}\lambda _{lm}\right) \\
-\lim \frac 1{4NT^{\prime }}\sum_{l=2} &&\sum_m\lambda _{lm}\left( \eta
(T^{\prime })\frac{4\pi }Q+\frac 1{2T^{\prime }}\lambda _{lm}\right) ^{-1}.
\end{eqnarray*}

In comparing the free energy per site on both sides of $T_c^{^{\prime }},$
we use the fact that the saddle point parameter is stuck at the value $\eta
^{*}$ for all $T_c^{^{\prime }}\leq T^{\prime }<0,$ that is, 
\[
\eta (T^{\prime })=\eta ^{*}\equiv -\frac Q{16\pi T_c^{^{\prime }}}, 
\]
to compute 
\begin{eqnarray*}
f(T_c^{^{\prime }} &<&T^{\prime })=\frac Q{16\pi }+\lim_{N\rightarrow \infty
}\frac{T^{\prime }}{2N}\sum_{l=2}\sum_m\ln \frac N{2T^{\prime }}\left(
-\frac 12+\lambda _{lm}\right) \\
f(T_c^{^{+}}) &=&\frac Q{16\pi }+\lim_{N\rightarrow \infty }\frac{%
T_c^{\prime }}{2N}\sum_{l=2}\sum_m\ln \frac N{2T_c^{\prime }}\left( -\frac
12+\lambda _{lm}\right) \\
f(T^{^{\prime }} &<&T_c^{\prime })=-\eta (T^{\prime })T^{\prime }+\lim \frac{%
T^{\prime }}{2N}\sum_{l=2}\sum_m\ln \left( N\eta (T^{\prime })\frac{4\pi }%
Q+\frac N{2T^{\prime }}\lambda _{lm}\right) \\
f(T_c^{-}) &=&\frac Q{16\pi }+\lim_{N\rightarrow \infty }\frac{T_c^{\prime }%
}{2N}\sum_{l=2}\sum_m\ln \frac N{2T_c^{^{\prime }}}\left( -\frac 12+\lambda
_{lm}\right)
\end{eqnarray*}
where equation (\ref{sadd1}) yields the saddle point parameter in terms of
temperature, $\eta =\eta (T^{\prime }).$ We deduce that the free energy per
site is the same on both sides of the critical point. This means there is no
latent heat involved in the phase transition at $T_c^{\prime }$ $<0$ and it
is therefore, a second order transition.

Since 
\begin{eqnarray*}
\left[ \frac d{dT}f\right] (T_c &<&T^{\prime })=-1+\lim_{N\rightarrow \infty
}\frac 1{2N}\sum_{l=2}\sum_m\ln \frac N{2T^{\prime }}\left( -\frac
12+\lambda _{lm}\right) \\
&=&-\frac{4\pi T_c^{\prime }}{QN}\sum_{l=2}^{\sqrt{N}}\sum_{m=-l}^l\left(
\lambda _{lm}-\frac 12\right) ^{-1}+K(T_c^{\prime }<T^{\prime }) \\
\left[ \frac d{dT}f\right] (T^{\prime } &<&T_c^{\prime })=-\eta (T^{\prime
})+\lim \frac 1{2N}\sum_{l=2}\sum_m\ln \frac N{2T^{\prime }}\left( \eta
(T^{\prime })T^{\prime }\frac{8\pi }Q+\lambda _{lm}\right) \\
-\lim_{N\rightarrow \infty } &&\frac 1{4NT^{\prime }}\sum_{l=2}\sum_m\lambda
_{lm}\left( \eta (T^{\prime })\frac{4\pi }Q+\frac 1{2T^{\prime }}\lambda
_{lm}\right) ^{-1} \\
&=&-\eta (T^{\prime })+K(T^{\prime }<T_c^{\prime })-L(T^{\prime
}<T_c^{^{\prime }}),
\end{eqnarray*}
and 
\[
T_c^{\prime }=\left[ \frac{4\pi }{QN}\sum_{l=2}^{\sqrt{N}}\sum_{m=-l}^l%
\left( \lambda _{lm}-\frac 12\right) ^{-1}\right] ^{-1}, 
\]
the above expressions for the derivative $\frac d{dT}f$ at both sides of the
critical temperature $T_c$ become 
\begin{eqnarray*}
\left[ \frac d{dT}f\right] (T_c^{+}) &=&-\lim_{N\rightarrow \infty }\frac{%
16\pi ^2T_c^{\prime }}{Q4\pi N}\sum_{l=2}^{\sqrt{N}}\sum_{m=-l}^l\left(
\lambda _{lm}-\frac 12\right) ^{-1}+K(T_c^{^{\prime }}) \\
&=&-\frac{16\pi ^2T_c^{\prime }}Q\left[ \lim_{N\rightarrow \infty }\frac
1{4\pi N}\sum_{l=2}^{\sqrt{N}}\sum_{m=-l}^l\left( \lambda _{lm}-\frac
12\right) ^{-1}\right] +K(T_c^{^{\prime }}) \\
\left[ \frac d{dT}f\right] (T_c^{-}) &=&\frac Q{16\pi T_c^{^{\prime
}}}+K(T_c^{^{\prime }})-L(T_c^{\prime }) \\
&=&\left[ \lim_{N\rightarrow \infty }\frac 1{4\pi N}\sum_{l=2}^{\sqrt{N}%
}\sum_{m=-l}^l\left( \lambda _{lm}-\frac 12\right) ^{-1}\right]
+K(T_c^{^{\prime }})-L(T_c^{\prime })
\end{eqnarray*}
where 
\[
0<K(T_c^{^{\prime }})\equiv \lim_{N\rightarrow \infty }\frac
1{2N}\sum_{l=2}\sum_m\ln \frac N{2T_c^{^{\prime }}}\left( -\frac 12+\lambda
_{lm}\right) <\infty 
\]
\[
-\infty <L(T_c^{^{\prime }})\equiv \lim_{N\rightarrow \infty }\frac
1{2N}\sum_{l=2}\sum_m\left( -\frac 12+\lambda _{lm}\right) ^{-1}\lambda
_{lm}<0 
\]
since 
\[
-\infty <\lim_{N\rightarrow \infty }\frac 1N\sum_{l=2}^{\sqrt{N}%
}\sum_{m=-l}^l\left( \lambda _{lm}-\frac 12\right) ^{-1}=\frac Q{4\pi T_c}<0 
\]
and for all $N,$ 
\[
\frac 1N\sum_{l=2}^{\sqrt{N}}\sum_{m=-l}^l\left( \lambda _{lm}-\frac
12\right) ^{-1}<\frac 1{2N}\sum_{l=2}\sum_m\left( -\frac 12+\lambda
_{lm}\right) ^{-1}\lambda _{lm}<0. 
\]

The significant point here is that the specific heat has a discontinuity at $%
T_c^{^{\prime }}:$%
\begin{eqnarray*}
\Delta &=&\left[ \frac d{dT}f\right] (T_c^{+})-\left[ \frac d{dT}f\right]
(T_c^{-}) \\
&=&L(T_c^{\prime })-\left( \frac{16\pi ^2T_c^{\prime }}Q+1\right) \left[
\lim_{N\rightarrow \infty }\frac 1{4\pi N}\sum_{l=2}^{\sqrt{N}%
}\sum_{m=-l}^l\left( \lambda _{lm}-\frac 12\right) ^{-1}\right] \\
&=&\lim_{N\rightarrow \infty }\frac 1{2N}\sum_{l=2}\sum_m\left( -\frac
12+\lambda _{lm}\right) ^{-1}\left( \lambda _{lm}-\left( \frac{8\pi
T_c^{\prime }}Q+\frac 1{2\pi }\right) \right)
\end{eqnarray*}
\begin{eqnarray*}
\frac Q{16\pi T_c^{\prime }} &=&\lim_{N\rightarrow \infty }\frac 1{4\pi
N}\sum_{l=2}^{\sqrt{N}}\sum_{m=-l}^l\left( -\frac 12+\lambda _{lm}\right)
^{-1}\simeq \frac 1{2\pi }\lim_{N\rightarrow \infty }\frac 1{2N}\sum_{l=2}^{%
\sqrt{N}}\frac{(2l+1)}{-1/2+1/l(l+1)} \\
&=&-\frac 1{2\pi }\lim_{N\rightarrow \infty }\frac 1N\sum_{l=2}^{\sqrt{N}}%
\frac{(2l+1)l(l+1)}{l(l+1)-2}>-\frac 1{2\pi }\lim_{N\rightarrow \infty
}\frac 1N\sum_{l=2}^{\sqrt{N}}(2l+1)>-\infty
\end{eqnarray*}
which is independent of enstrophy $Q$ since $L(T_c^{\prime })$ is and $%
T_c^{\prime }$ is proportional to $Q,$ and which we expect to be positive.

From the expression 
\[
f(T_c^{^{\prime }}<T^{\prime })=\frac Q{16\pi }+K(T^{\prime })T^{\prime } 
\]
we deduce that $f(T_c^{^{\prime }}<T^{\prime })$ increases as $T^{\prime }$
increases away from $T_c^{^{\prime }}$ and also that $K(T^{\prime
})T^{\prime }$ consists of the sum 
\[
-T^{\prime }s(T^{\prime })+\left( u(T^{\prime })-\frac Q{16\pi }\right) 
\]
where $s(T^{\prime })$ is the entropy per site and $u(T^{\prime })$ is the
internal energy per site. At $T^{\prime }=0,$ the internal energy per site $%
u(T^{\prime })=\frac Q{16\pi }$ consists entirely of energy in the ground
modes. At $T_c^{^{\prime }}<T^{\prime }<0,$ this represents the fact that $%
u(T^{\prime })-\frac Q{16\pi }>0$ is that part of the internal energy in the
ergodic modes.

From (\ref{sadd1}), we deduce that $\eta =\eta (T^{\prime })$ decreases as $%
T^{\prime }$ becomes more negative than $T_c^{^{\prime }}.$ Then, from 
\[
f(T^{^{\prime }}<T_c^{\prime })=-\eta T^{\prime }+\lim \frac{T^{\prime }}{2N}%
\sum_{l=2}\sum_m\ln \left( N\eta \frac{4\pi }Q+\frac N{2T^{\prime }}\lambda
_{lm}\right) 
\]
we deduce that $f(T^{^{\prime }}<T_c^{\prime })$ decreases as $T^{\prime }$
decreases away from $T_c^{^{\prime }}.$ Here, all the internal energy is in
the ergodic modes and none in the ground modes, and the entropic term $%
-T^{\prime }s(T^{\prime })$ becomes more dominant as $T^{\prime }<0$
decreases.

\end{document}